\documentclass[aps,prd,amssymb,eqsecnum,nofootinbib,a4paper]{revtex4}   
\newif\ifusesec
\usesectrue 
\usepackage{geometry}                
\usepackage{graphics}    
\usepackage{epstopdf}
\usepackage{helv et,amsmath,amssymb,amsfonts,bm} 
\usepackage[latin1]{inputenc} 
\usepackage[T1]{fontenc} 
\usepackage[english]{babel}
\usepackage{graphicx,epsfig,stackrel} 
\usepackage[all]{xy}
\usepackage{hyperref}

\newcommand{\be}{\begin{equation}}
\newcommand{\ee}{\end{equation}}

\begin{document}
\title{Quantum Supersymmetric Cosmological Billiards \\ and their Hidden Kac-Moody Structure}

\author{Thibault Damour}
\email{damour@ihes.fr}
\affiliation{Institut des Hautes Etudes Scientifiques, 35 route de Chartres, 91440 Bures-sur-Yvette, France}

\author{Philippe Spindel}
\email{philippe.spindel@umons.ac.be}
\affiliation{Unit\'e de  M\'ecanique et Gravitation, Universit\'e de Mons, Facult\'e des Sciences,20, Place du Parc, B-7000 Mons, Belgium}
\date{\today}

\begin{abstract}
We study the  quantum fermionic billiard defined by the dynamics of a quantized supersymmetric squashed three-sphere
(Bianchi IX cosmological model within $D=4$ simple supergravity). The quantization of the homogeneous gravitino field leads to
a 64-dimensional fermionic Hilbert space. We focus on the 15- and 20-dimensional subspaces (with fermion numbers $N_F=2$
and $N_F=3$) where there exist propagating solutions of the supersymmetry constraints that carry (in the small-wavelength limit)
a chaotic spinorial dynamics  generalizing the Belinskii-Khalatnikov-Lifshitz  classical ``oscillatory" dynamics. By exactly solving the supersymmetry
constraints near each one of the three dominant potential walls underlying the latter chaotic billiard dynamics, 
we compute the three operators that describe the corresponding three potential-wall reflections  of the spinorial state describing,
in supergravity, the quantum evolution of the universe. It is remarkably found that the latter, purely dynamically-defined, reflection operators
satisfy generalized Coxeter relations which define a type of spinorial extension of the Weyl group of the rank-3 hyperbolic Kac-Moody
algebra $AE_3$.
\end{abstract}
\maketitle
\section{Introduction}

One of the challenges of gravitational physics is to describe the fate of spacetime at spacelike singularities (such as
the cosmological big bang, or big crunches within black holes).  
A new avenue for attacking this problem has been suggested a few years ago via a conjectured {\it correspondence} between various supergravity theories and the dynamics of a spinning massless particle on an infinite-dimensional Kac-Moody coset space~\cite{Damour:2002cu,Damour:2005zs,de Buyl:2005mt,Damour:2006xu}. 
Evidence for such a supergravity/Kac-Moody link emerged through the study \`a la Belinskii-Khalatnikov-Lifshitz (BKL)~\cite{Belinsky:1970ew} of the structure of cosmological singularities in string theory and supergravity, in spacetime dimensions $4 \leq D \leq 11$~\cite{Damour:2000hv,Damour:2001sa,Damour:2002et}. [For a different approach to such a conjectured supergravity/Kac-Moody link see
\cite{West:2001as,Tumanov:2016dxc}.]
For instance,  the well-known BKL oscillatory behavior~\cite{Belinsky:1970ew}  of the diagonal components of a generic, inhomogeneous Einsteinian metric in $D=4$ was found to be equivalent to a billiard motion within the Weyl chamber of the rank-3 hyperbolic Kac-Moody algebra $AE_3$~\cite{Damour:2001sa}. Similarly,
the generic BKL-like dynamics of the bosonic sector of maximal supergravity (considered either in $D=11$, or, after dimensional reduction, in $4 \leq D \leq 10$) leads to a chaotic billiard motion within the Weyl chamber of the rank-10 hyperbolic Kac-Moody algebra $E_{10}$~\cite{Damour:2000hv}.  The hidden r\^ole of $E_{10}$ in the dynamics of maximal supergravity was confirmed to higher-approximations (up to the third level) in the gradient expansion $\partial_x \ll \partial_T$ of  its bosonic sector~\cite{Damour:2002cu}. In addition, the  study of the fermionic sector of supergravity theories has exhibited a related r\^ole of Kac-Moody algebras. At leading order in the gradient expansion of the gravitino field $\psi_{\mu}$,  the dynamics of $\psi_{\mu}$ at each spatial point was found to be given by  parallel transport with respect to a (bosonic-induced) connection $Q$ taking values within the ``compact'' sub-algebra of the corresponding bosonic Kac-Moody algebra: say $K(AE_3)$ for $D=4$ simple supergravity and $K(E_{10})$ for maximal supergravity~\cite{Damour:2005zs,de Buyl:2005mt,Damour:2006xu}.  This led to the
study of fermionic cosmological billiards \cite{Damour:2009zc,Kleinschmidt:2009cv}. [For definitions, and basic mathematical 
results on Kac-Moody algebras see Ref. \cite{Kac}; see also Ref. \cite{FF} for a detailed study of the specific hyperbolic Kac-Moody algebra
$AE_3 \equiv {\mathcal F}$ that enters 4-dimensional gravity and supergravity.]

The works cited above considered only the terms {\it linear} in the gravitino, and, moreover, treated  $\psi_{\mu}$ as a ``classical'' (i.e. Grassman-valued) fermionic field. It is only recently  \cite{Damour:2013eua,Damour:2014cba} that the full quantum supergravity dynamics of simple cosmological models has been tackled in a way which displayed their hidden Kac-Moody structures. [For previous work on supersymmetric quantum cosmology, see Refs. \cite{D'Eath:1993up,D'Eath:1993ki,Csordas:1995kd,Csordas:1995qy,Graham:1995ni,Cheng:1994sr,Cheng:1996an,Obregon:1998hb}, as well as the books \cite{D'Eath:1996at,VargasMoniz:2010rxa}.]

The work \cite{Damour:2013eua, Damour:2014cba}  studied the quantum supersymmetric Bianchi IX cosmological model. This model is obtained by the (consistent) dimensional reduction of the simple $N=1$, $D=4$ supergravity to one (timelike) dimension on a triaxially-squashed 
($SU(2)$-homogeneous) three-sphere.  This work allowed to decipher the quantum dynamics of this supersymmetric (mini-superspace) model.
The quantum state  $| \Psi(\beta) \rangle$ of this model depends (after a symmetry reduction) on three continuous bosonic parameters $\beta^a$, $a=1,2,3$  (measuring the triaxial squashing of the three-sphere), and on sixty-four spinor indices (which describe the representation space of the anticommutation relations
of the gravitino field displayed below).
It was shown that the structure of the solutions of the supersymmetry (susy) constraints depended very much
on the eigenvalue $N_F$ (going from 0 to 6) of the fermion-number operator:
\begin{equation}\label{NF}
\widehat N_F = 3+\frac12 \, G_{ab} \, \overline{\widehat \Phi}^a \gamma^{\hat1\hat2\hat3} \, \widehat \Phi^{b}
\end{equation}
Here, ${\widehat \Phi}^{a}_A$ (with a spatial vector index $a=1,2,3$, and with a Majorana spinor index $A=1,2,3,4$ that we generally suppress) denote the twelve, quantized
homogeneous modes of the spatial components of the gravitino field $\psi_{\mu}$ (written in a special way that makes more manifest some of their Kac-Moody properties).
They satisfy the anticommutation relations
\be \label{clifford}
\widehat \Phi^{a}_A \, \widehat \Phi^{b}_B  + \widehat \Phi^{b}_B \, \widehat \Phi^{a}_A  =  G^{ab} \delta_{AB}
\ee
where 
\begin{equation}\label{Gabu}
 G^{ab}  =\frac 12 \begin{pmatrix}
1 &-1 &-1 \\
-1 &1 &-1 \\
-1 &-1 &1 \end{pmatrix} \qquad .
\end{equation}
defines a contravariant, Lorentzian-signature [$(-,+,+)$] metric in the three-dimensional space spanned by the bosonic variables $\beta^a$.
[See Ref. \cite{Damour:2014cba} for more details on our notation.]

The quantum state  $| \Psi(\beta) \rangle$ must be annihilated by the susy constraints, i.e.
\begin{equation} \label{susy}
\widehat {\cal S}^{(0)}_A \vert \Psi(\beta)\rangle = 0 \, ,
\end{equation}
where the structure of the susy constraints is
\begin{equation}\label{SA}
\widehat {\cal S}^{(0)}_A =  \frac{i}{2} \widehat \Phi^a_A \partial_{\beta^a} + \widehat V_A(\beta, \widehat \Phi)  \qquad .
\end{equation}
Here the potential-like term $\widehat V_A(\beta, \widehat \Phi)$ is a complicated operator which is {\it cubic} in the
gravitino operators $\widehat \Phi^a_A$, and involves various potential walls that will be discussed below.

As the twelve $\widehat \Phi^a_A$'s
satisfy the Clifford-algebra anticommutation law \eqref{clifford}, and as the $\widehat \Phi^a_A$'s enter the first term of 
$\widehat {\cal S}^{(0)}_A$, Eq. \eqref{SA}, as coefficients of the partial derivatives $\partial_{\beta^a}$,
 we can view, for each given value of the index $A$, the susy constraint \eqref{susy} as being a Diraclike 
 [$ i \gamma^{\mu}\partial_{x^{\mu}} \psi(x)= \widehat V(x) \psi(x)$] equation for the propagation
 of the wavefunction $| \Psi(\beta) \rangle$ in the 3-dimensional Lorentzian $\beta$ space.
 However, as the Majorana-spinor index $A$ in Eqs. \eqref{susy},  takes four values, we see that the state must
 simultaneously solve four different Diraclike equations. This represents a huge constraint on possible solutions.
 
 The structure of the solution space of these susy constraints has been thoroughly analyzed in \cite{Damour:2014cba}.
 It was found that the structure and generality of the solutions drastically depend on the fermionic level $N_F$, Eq. \eqref{NF}.
 Here, we shall study the cosmological dynamics of the solutions at levels\footnote{There are similar solutions at level $N_F=4$,
 and in the mirror part of the $N_F=2$ level that we shall not consider, which can be obtained by a simple involution acting on fermionic generators.}
 $N_F=2$ and $N_F=3$ that contain two arbitrary
 real functions of two variables as free Cauchy data, i.e. that have as much freedom as the solutions of the usual, purely bosonic 
 Bianchi IX mini-superspace Wheeler-DeWitt equation. More precisely, we are interested in quantum solutions which, in the
 WKB approximation, can be viewed as describing the chaotic billiard motion of the cosmological squashing parameters
 $\beta^1, \beta^2, \beta^3$ near a big-crunch-type singularity. [This chaotic behavior is a quantum, and spinorial,
 generalization of the classic BKL oscillatory behavior of the three Bianchi IX scale factors,
 $a=e^{-\beta^1}, b=e^{-\beta^2}, c==e^{-\beta^3}$. The quantum (scalar) version of the Bianchi IX chaos was
 first studied in Ref. \cite{Misner:1969ae}.] The type of solution we have in mind, and will study in detail
 below, is illustrated in Fig. 1.
 \begin{figure}
\includegraphics[scale=0.7]{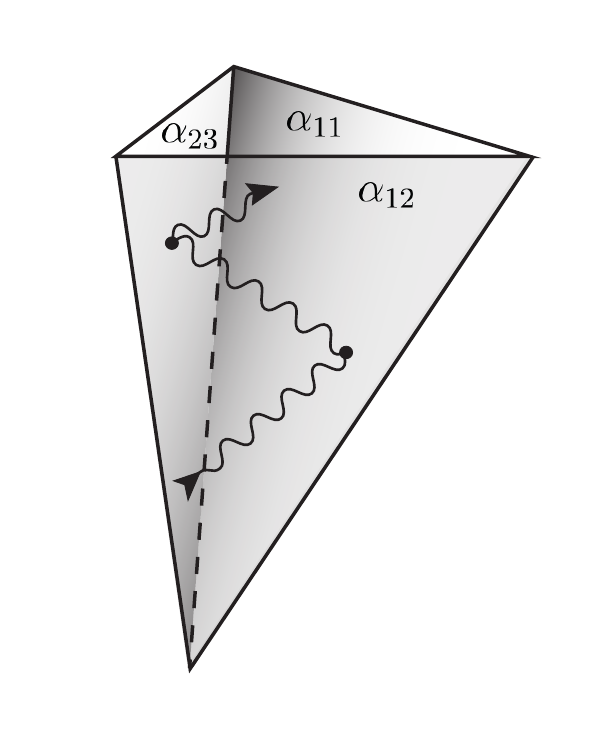}
\caption{Sketchy representation of the propagation in 3-dimensional, Lorentzian  $\beta$ space of the cosmological quantum supergravity
wave function $|\Psi(\beta)\rangle $. When considered within our canonical chamber $\beta^1<\beta^2<\beta^3$, this wave function undergoes
successive reflections  on the three potential walls that are present in the supersymmetry constraints \eqref{susy}.
Two of the potential walls are singular on the hyperplanes $\alpha_{12}(\beta)=0$ and $\alpha_{23}(\beta)=0$,
while the third potential wall grows  exponentially when  $\alpha_{11}(\beta)$ becomes negative.}
\end{figure}

 As illustrated on Fig. 1, we can view these solutions as wave packets bouncing between potential walls. In Fig. 1, these
 potential walls are drawn as sharp walls located on some (timelike) hyperplanes in $\beta$-space. [Note, however, that our
 analysis will not make any sharp-wall approximation, as was made, e.g., in Ref. \cite{Kleinschmidt:2009cv}. We will compute the
 reflection of the wave function against each exact potential wall; see below.]
 In particular, we highlighted the
 three wall hyperplanes defined by the equations 
 \be \label{chamber}
 \alpha_{11}(\beta)=0 \, , \, \alpha_{12}(\beta)=0 \,, \, \alpha_{23}(\beta)=0 ,
 \ee
 corresponding to the following three linear forms in the $\beta$'s:
 \be\label{defaij}
 \alpha_{11}(\beta)\equiv 2 \beta^1 \,;\, \alpha_{12}(\beta)\equiv\beta^2-\beta^1\, ; \, \alpha_{23}(\beta)\equiv\beta^3-\beta^2 \,.
 \ee
The three hyperplane equations \eqref{chamber} constitute a conventional way of describing the fact that the basic
equations of the supersymmetric Bianchi IX model, i.e. the susy constraints \eqref{susy}, contain operatorial, spin-dependent
and $\beta$-dependent potentiallike terms that grow when the $\beta$'s approach these hyperplanes. More precisely,
as can be seen on the explicit expressions given in Eqs. (6.1)--(6.4) of \cite{Damour:2014cba}, 
the potential-like contribution $\widehat V_A(\beta, \widehat \Phi)$ to the susy constraints
operators, Eq. \eqref{SA}, contains the following terms
\begin{equation}
\label{Vg}
\widehat {V}_A^g = \frac12 \sum_a e^{-2\beta^a} \left(\gamma^5 \, \widehat \Phi^a\right)_A\qquad ,
\end{equation}
and
\begin{eqnarray}
\label{Vs}
\widehat {V}_A^{\rm sym} &= &-\frac18 \coth [\beta^1 - \beta^2] \left[ \widehat S_{12} \left( \gamma^{\hat1\hat2} \left( \widehat \Phi^1 - \widehat \Phi^2 \right)\right)_A + \left( \gamma^{\hat1\hat2} \left( \widehat \Phi^1 - \widehat \Phi^2 \right)\right)_A \widehat S_{12} \right] \nonumber \\
&+&{\rm cyclic}_{123}
\end{eqnarray}
where
\begin{equation}
\label{Spin}
\widehat S_{12} =\frac12 \left( \overline{\widehat\Phi}^3 \, \gamma^{\hat 0\hat 1\hat 2} (\widehat\Phi^1 + \widehat\Phi^2) + \overline{\widehat\Phi}^1 \, \gamma^{\hat 0\hat 1\hat 2} \, \widehat\Phi^1
+ \overline{\widehat\Phi}^2 \, \gamma^{\hat 0\hat 1\hat 2} \, \widehat\Phi^2 - \overline{\widehat\Phi}^1 \, \gamma^{\hat 0\hat 1\hat 2} \, \widehat\Phi^2 \right)   .
\end{equation}
The operator  $ \widehat S_{12}$, together with  similarly defined operators $\widehat S_{23},\ \widehat S_{31}$,  are spin-like operators satisfying the usual $su(2)$ commutation relations : $[\widehat S_{23},\ \widehat S_{31}]=+ i\,\widehat S_{12}$, etc . 

The ``gravitational-wall" potential term \eqref{Vg} is exponentially small  when $\beta^1$, $\beta^2$, and $\beta^3$, are largish
and positive. It starts becoming exponentially large (and confining) when, on the contrary, either $\beta^1$, $\beta^2$, or $\beta^3$,
become negative. It is in that sense that the three gravitational-wall hyperplanes   $\alpha_{11}(\beta)=0$,  $\alpha_{22}(\beta)=0$
and  $\alpha_{33}(\beta)=0$  (where $\alpha_{22}\equiv 2 \beta^2$ and $\alpha_{33}\equiv 2 \beta^3$) define (softly confining)
potential walls.
The ``symmetry-wall" potential \eqref{Vs} is similarly made of three different terms (differing by  a $(123)$ cyclic permutation).
For instance, the term explicitly displayed in Eq. \eqref{Vs}, which involves $\coth [\beta^1 - \beta^2] $, is singular on
the symmetry hyperplane $\alpha_{12}(\beta)=0$, and tends towards a $\beta$-independent contribution far from it.
[The various $\beta$-independent contributions coming from the asymptotic $\pm 1$ values of the various $\coth \alpha_{ab}$'s
combine with other $\beta$-independent, $\widehat \Phi^a$-cubic terms to define an effective mass term in the above
Diraclike equations. The effect of these mass-like, $\widehat \Phi^a$-cubic terms will be fully taken into account in our discussion below.]

It has been shown in \cite{Damour:2014cba} that it is enough to consider the evolution of the universe wave function
$|\Psi(\beta)\rangle $ within only one of the six different chambers defined by considering the two possible sides associated with
 the three symmetry-wall one forms
$\alpha_{12}(\beta)$, $\alpha_{23}(\beta)$, $\alpha_{31}(\beta)$ (i.e. the two possible signs for, e.g., $\beta^2-\beta^1$).
Each such chamber corresponds to some ordering of the three $\beta$'s. Here, we shall work within the canonical chamber
\be\label{chamber0}
\beta^1< \beta^2 <\beta^3  \quad .
\ee
The gravitational wall belonging to this chamber [namely the term $e^{-2\beta^1} \left(\gamma^5 \, \widehat \Phi^1\right)_A$
in \eqref{Vg}] further confines the evolution of the wave packet to stay essentially on the positive side of $\alpha_{11}= 2 \beta^1$,
so that we can think of the wave function $|\Psi(\beta)\rangle $ as evolving in the (approximate) billiard chamber
\be \label{chamber1}
0 \lesssim \beta^1< \beta^2 <\beta^3 \quad .
\ee
It is this (approximate) billiard chamber, within which  $ \alpha_{11}(\beta) \geq 0 \, , \, \alpha_{12}(\beta) \geq0 \,, \, \alpha_{23}(\beta)\geq 0$,   which is represented in Fig. 1.

In the present work, we shall complete the results of Refs. \cite{Damour:2013eua,Damour:2014cba} by studying the {\it quantum}
reflection operators of the universe wave function $|\Psi(\beta)\rangle $ on the three potential walls \eqref{chamber} constraining
its propagation (see Fig. 1). Our present study will thereby represent the quantum generalization of Ref. \cite{Damour:2009zc}, which studied
similar reflection operators when treating the gravitino as a classical, i.e. Grassmann variable. We will study, in turn, the 
evolution of $|\Psi(\beta)\rangle $ at the fermionic level $N_F=2$ (Sec. II) and at the fermionic level $N_F=3$ (Sec. III).
After the completion of this purely dynamical problem, we shall show (in Sec. IV) that our results provide a new evidence
for the hidden role of the hyperbolic Kac-Moody algebra $AE_3$ (and of its compact subalgebra $K[AE_3]$) in supergravity.

\section{Quantum fermionic billiard at level $N_F = 2$}\label{SW615}

The susy constraints, Eqs. (\ref{susy}), admit solutions depending on arbitrary functions at level $N_F=2$  only in a 6-dimensional subspace  
of the total 15-dimensional $N_F=2$ space, namely  (with $p,q=1,2,3$):
\begin{equation}\label{kpq615}
|\Psi \rangle_{{\bf 6}, N_F=2}= k_{pq}(\beta) \tilde b_+^{(p} \tilde b_-^{q)} \, \vert 0 \rangle_-   \quad .
\end{equation}
Here, the amplitude $k_{pq}(\beta)$ parametrizing these solutions is symmetric in the two indices $p,q=1,2,3$,
and the two triplets of operators $\tilde b^a_\pm = ( b^a_\pm )^\dagger$ denote the Hermitian conjugates of the following combinations of the 
basic (Hermitian) gravitino operators $\widehat\Phi^a_A$
\be
b_+^a = \widehat\Phi^a_1 + i \widehat\Phi^a_2 \,  ; \, b_-^a = \widehat\Phi^a_3 - i \widehat\Phi^a_4 \quad .
\ee
The vacuum state $\vert 0 \rangle_-$ is the unique state annihilated by the six fermionic annihilation operators $ b^a_\pm$.
The  total 15-dimensional $N_F=2$ space is generated by acting  on $\vert 0 \rangle_-$ with two among the six (anticommuting) creation operators $\tilde b^a_\pm$.
The generic propagating state \eqref{kpq615} lives in the 6-dimensional subspace ${\mathbb H}_{(1,1)_S}$ 
spanned by the symmetrized products $\tilde b_+^{(p} \tilde b_-^{q)} \, \vert 0 \rangle_-$.

In this Section we shall discuss the reflection law of the $N_F=2$ spinorial solutions \eqref{kpq615} against the three different
potential walls bounding the  chamber within which these solutions propagate. We are interested in an
asymptotic regime (large $\beta$'s, and small wavelengths) where the quantum solutions can be approximated (away from
the turning points, i.e. sufficiently away from the potential walls) by quasi-classical WKB solutions (see Fig. 1). 
Like in the usual WKB approximation, 
we will obtain the reflection laws against the potential walls by matching the WKB form (away from the walls)
to  (exact)  solutions valid near the walls.

Far from all the walls (in our canonical Weyl chamber $ 0 \lesssim \beta^1 \leq \beta^2 \leq \beta^3$),  the effect of the
$\beta$-dependent potential terms is negligible, so that the amplitude $k_{pq}(\beta)$ of the general WKB-like
$N_F=2$ spinorial solution can be written as superposition of (rescaled) plane waves:
\begin{equation} \label{NF2wave}
k_{pq}^{\rm far-wall}(\beta) = F(\beta)  \sum \, K_{pq} \, e^{i\pi'_a\beta^a} \,.
\end{equation}
Here, the rescaling factor $F(\beta)$ is generally defined (see Eq. (8.4) in Ref. \cite{Damour:2014cba}, here modified by the numerical
factor $8^{-1/8}$) as
\begin{equation} \label{F}
 F(\beta)  = e^{\frac34 \, \beta^0} (8 | \sinh \beta_{12} \, \sinh \beta_{23} \, \sinh \beta_{31}| )^{-1/8} \, ,
  \end{equation} 
 where we introduced the convenient short-hands
 \be
 \beta^0 \equiv   \beta^1 +  \beta^2 +  \beta^3\ ,\  \beta_{12}\equiv\beta^1-\beta^2\ ,\ \text{etc}\qquad.
 \ee
Far from all the walls of the canonical chamber, the  rescaling factor $F(\beta)$ is a (real) exponential of the $\beta$'s,
namely
\be
F(\beta) \approx e^{\frac34 \, \beta^0} e^{-\frac18 (|\beta_{12}| + |\beta_{23}| + |\beta_{31}|)}= e^{\beta^1+\frac 34\beta^2+\frac 12\beta^3} \,.
\ee
As explained in Refs. \cite{Damour:2013eua,Damour:2014cba}, the rescaling factor $F(\beta)$ is such that the mass-shell condition for the plane wave factor $ e^{i\pi'_a\beta^a}$
takes the simple, special-relativistic-like, form $ \pi'^2= - \mu^2$, namely
\be
 G^{ab} \pi'_a  \pi'_b= -\mu^2_{N_F=2}= +\frac38 \,,
\ee
where $G^{ab}$ denotes the Lorentzian-signature (inverse) metric in $\beta$-space. Note that the $N_F=2$ mass-shell
is tachyonic ($ \mu^2_{N_F=2}= - \frac38$, i.e. $\pi'_a$ is a spacelike momentum).
 As was discussed in Ref. \cite{Damour:2014cba}, this tachyonic character (which holds for all
fermionic levels, except $N_F=3$) suggests the possibility of a cosmological bounce. In the present study, we are, however,
focussing on an intermediate asymptotic regime where the wavepacket is centered, most of the time, around coordinates $\beta^a$
that are large compared to 1, so that many wavelengths separate the successive wall reflections.

The amplitude $K_{pq}$ (a ``tensor" in $\beta$-space) of each plane wave in Eq. \eqref{NF2wave} was found in Ref. \cite{Damour:2014cba}
to have (for a given momentum vector $\pi'_a$) only one (complex) degree of freedom, contained in
an overall factor, say $C_{N_F=2}$, i.e. to be of the form
\begin{equation}
\label{farwall}
K_{pq} = C_{N_F=2}\left(\pi'_p \, \pi'_q + L_{pq}^k \, \pi'_k + m_{pq}\right) \,,
\end{equation}
where $L_{pq}^k$ and  $m_{pq}$ are some fixed numerical coefficients (see Eqs. (19.17) and (19.18) in 
\cite{Damour:2014cba}, which are reproduced in Appendix A for the reader's convenience).

A first way of  describing the law of reflection of a plane wave \eqref{NF2wave} on a potential wall is to compute the
transformation between the incident values of  the overall amplitude  and of the momentum, say $C_{N_F=2}^{\rm in}$, ${\pi'}_a^{\rm in}$,
and their reflected (or outgoing) values, say $C_{N_F=2}^{\rm out}$, ${\pi'}_a^{\rm out}$.  In order to derive the scattering map
$C_{N_F=2}^{\rm in} \to C_{N_F=2}^{\rm out}$, ${\pi'}_a^{\rm in} \to {\pi'}_a^{\rm out}$ we need to go beyond the far-wall approximation,
and study the behaviour of a generic wave packet \eqref{NF2wave} near each type of potential wall.

Anticipating on the results of the computations given in the following subsections, let us already exhibit the simple structure of
the scattering maps. The transformation of the momentum $\pi'_a$ upon reflection on a potential wall associated with a root 
$\alpha(\beta)$ is simply given (as expected from the classical billiard approximation) by specular reflection (with respect to
the $\beta$-space geometry defined by the (contravariant) metric $G^{ab}$), i.e. by
\be \label{reflection1}
{\pi'}_a^{\rm out} = {\pi'}_a^{\rm in} - 2 \frac{\pi'^{\rm in} \cdot \alpha}{\alpha \cdot \alpha} \, \alpha_a \,.
\ee
Here, the scalar product between two covariant vectors is defined by $ \pi' \cdot \alpha \equiv  G^{ab} \pi'_a \alpha_b$.
[$\alpha_a$ is the covariant normal to the considered potential wall, which is ``located" on the hypersurface 
$0=\alpha(\beta) \equiv \alpha_a \beta^a$.] 

As for the transformation of the overall scalar amplitude $C_{N_F=2}$, it will be found to be encoded in a global phase,
$\delta_{\alpha}^{\rm global}$ (which will depend on the considered type of potential wall):
\be\label{reflection2}
C_{N_F=2}^{\rm out} = e^{i \delta_{\alpha}^{\rm global}}  \, C_{N_F=2}^{\rm in}  \,.
\ee

A second way of describing the law of reflection of a plane wave \eqref{NF2wave} on a potential wall $\alpha$ is to compute the
``reflection operator" $\mathcal{ R}_\alpha$, acting on the Hilbert space where the considered quantum spinorial state lives,
and transforming the incident state $|\Psi \rangle^{\rm in}$ into the corresponding reflected state $|\Psi \rangle^{\rm out}$.
In the present case, the considered Hilbert space is the 6-dimensional subspace ${\mathbb H}_{(1,1)_S}$ of the 15-dimensional
$N_F=2$ level, and the incident state is the ingoing part of  \eqref{kpq615}, i.e. a plane-wave state of the type
$F(\beta)  \, K_{pq} \, e^{i {\pi'}^{\rm in}_a\beta^a} \tilde b_+^{(p} \tilde b_-^{q)} \, \vert 0 \rangle_-$.
The corresponding reflection operator then acts on ${\mathbb H}_{(1,1)_S}$ and is such that
\be\label{reflection3}
|\Psi \rangle^{\rm out}_{{\bf 6}, N_F=2}= \mathcal{ R}_\alpha^{{\bf 6}, N_F=2}   |\Psi \rangle^{\rm in}_{{\bf 6}, N_F=2} \quad .
\ee
[When considering the action of $\mathcal{ R}_\alpha$ we strip $|\Psi \rangle^{\rm in}$ and $|\Psi \rangle^{\rm out}$
of their corresponding phase factors $e^{i {\pi'}^{\rm in/out}_a\beta^a}$.]
As the fundamental billiard chamber of the supersymmetric Bianchi IX model is bounded by three walls, described by three
linear forms in $\beta$-space, namely $\alpha_{12}(\beta) = \beta^2- \beta^1, \alpha_{23}(\beta) = \beta^3- \beta^2$ 
and $\alpha_{11}(\beta) = 2 \beta^1$, the quantum supersymmetric 
Bianchi IX billiard will define (at each fermionic level where there exists propagating states) three different reflection
operators. For instance, at the  $N_F=2$ level, supergravity will define three spinorial reflection operators
\be\label{reflection4}
  \mathcal{ R}_{\alpha_{12}}^{{\bf 6}, N_F=2},  \mathcal{ R}_{\alpha_{23}}^{{\bf 6}, N_F=2},  \mathcal{ R}_{\alpha_{11}}^{{\bf 6}, N_F=2}, \ee
  all acting in the same 6-dimensional space ${\mathbb H}_{(1,1)_S}$. We shall compute these (dynamically defined) operators below,
and find that they have a remarkable Kac-Moody meaning.

In order to derive the reflection laws \eqref{reflection1}, \eqref{reflection2}, \eqref{reflection3},  and,
in particular, to compute the values of the global phases $\delta_{\alpha}^{\rm global}$, and of the reflection operators, \eqref{reflection4},
we will use a ``one-wall"  approximation, i.e. we shall separately solve the problems 
where an asymptotically planar wave $ F(\beta)  \, K_{pq} \, e^{i\pi'_a\beta^a}$ impinges on one of the
three possible walls of our canonical chamber, $ 0 \lesssim \beta^1 \leq \beta^2 \leq \beta^3$, i.e. either 
on one of the two symmetry walls $\alpha_{12}(\beta) = \beta^2- \beta^1$, or $\alpha_{23}(\beta) = \beta^3- \beta^2$;
or on the gravitational wall $\alpha_{11}(\beta) = 2 \beta^1$. In this one-wall approximation, the spinorial wavefunction
$k_{pq}(\beta)$ in \eqref{kpq615} will essentially depend only on one variable (measuring the orthogonal distance to the wall),
which will make the problem of exactly solving the complicated supersymmetry constraints \eqref{susy} tractable. 

\subsection{Scattering on the symmetry wall $\alpha_{23}(\beta) = \beta^3- \beta^2$.}

In this subsection we study (in the one-wall approximation) the reflection of the $N_F=2$ spinorial state \eqref{kpq615}
on the symmetry wall $\alpha_{23}(\beta) = \beta^3- \beta^2$. This study is simplified by using an adapted basis in
$\beta$-space. In doing so, we shall treat the building blocks entering \eqref{kpq615} as tensors, with the indicated
variance, in $\beta$-space. Namely, each creation operator  $\tilde b_{\pm}^{p}$ is considered as a (contravariant)
vector, while the amplitude $  k_{pq}$ is viewed as a (symmetric) covariant 2-tensor.  Given a basis of 1-forms (i.e. a set of three
independent linear forms in $\beta$-space), say $\alpha^{\widehat 1}(\beta)= \alpha^{\widehat 1}_p  \, \beta^p$,
$\alpha^{\widehat 2}(\beta)= \alpha^{\widehat 2}_p  \, \beta^p$, $\alpha^{\widehat 3}(\beta)= \alpha^{\widehat 3}_p  \, \beta^p$,
we shall then work with the corresponding basis (or dual basis) components $\tilde b_{\pm}^{\widehat a}\equiv \alpha^{\widehat a}_p \, \tilde b_{\pm}^{p}$
and $ k_{\widehat a \widehat b}\equiv \alpha_{\widehat a}^p \, \alpha_{\widehat b}^q \,  k_{p q}$, where we defined $\alpha_{\widehat a}^p \alpha^{\widehat b}_p\equiv \delta_{\widehat a}^{\widehat b}$.

It is very useful to use a basis of the type
\be \label{basisNF2a23}
\left\{\alpha^{\perp}, \alpha^u, \alpha^v \right\} \equiv \left\{\alpha(\beta), u(\beta), v(\beta)\right\} \,,
\ee
where $\alpha(\beta)$ is the reflecting wall form we are considering, i.e., in the present subsection
\be \label{a23}
\alpha(\beta) \equiv \alpha^{\perp}(\beta) \equiv \alpha_{23}(\beta) = \beta^3- \beta^2 ,
\ee
while $u(\beta), v(\beta)$ are two one-forms whose corresponding contravariant vectors\footnote{The triplet of contravariant
vectors  $\left\{ \alpha^{\sharp},  u^{\sharp},  v^{\sharp} \right\}$ should not be confused with the vectorial
basis that is {\it dual} to the basis of one-forms  \eqref{basisNF2a23}. As we shall see below
the dual basis $\left\{\alpha_{\perp} , \alpha_u, \alpha_v \right\}$ is $\left\{ \frac12 \alpha^{\sharp}, \frac12 v^{\sharp}, \frac12 u^{\sharp} \right\}$.}
(with $u^{\sharp \, p} \equiv G^{p q} u^{\sharp}_q$), are {\it parallel} to the wall hyperplane $\alpha(\beta)=0$, i.e. $ \alpha^{\perp} (u^{\sharp}) =0 = \alpha^{\perp} (v^{\sharp})$. Geometrically, the (contravariant) vector $\alpha^{\sharp}$ is perpendicular
to the wall hyperplane $\alpha(\beta)=0$, while the numerical function $\beta \to \alpha(\beta)$ measures (modulo a factor $\sqrt{2}$) the orthogonal distance
away from the wall hyperplane. [The squared norms of the wall forms we shall consider here are all equal to 2: $ \alpha \cdot \alpha = G^{pq} \alpha_p \alpha_q = 2$. This normalization is adapted to the Kac-Moody interpretation of the (dominant) wall forms as simple roots of
a Kac-Moody Lie algebra.]

It was further found to be convenient to align the two basis elements which are parallel to the wall to the two (intrinsically defined) 
{\it null} directions tangent to the wall. [The wall hyperplane is spacelike in Lorentzian $\beta$-space, so that it intersects the lightcone 
$G_{pq} \beta^p \beta^q=0$ along two lines.] Specifically, we use
\begin{eqnarray}\label{uv}
u(\beta)&\equiv&-(2\,\beta^1+\frac 12\,\beta^2+\frac 12\,\beta^3)\qquad ,\\
v(\beta)&\equiv&\beta^2+\beta^3\qquad .
\end{eqnarray}
The only nonzero scalar products among the three basis one-forms $\left\{\alpha^{\perp}, \alpha^u, \alpha^v \right\} \equiv \left\{\alpha(\beta), u(\beta), v(\beta)\right\}$ are $ \alpha \cdot \alpha = 2$ and $ u \cdot v = 2$, so that the only nonzero components of the
inverse metric $G^{\widehat a  \, \widehat b}$ are
\be \label{Gup23}
G^{\perp \perp} =  G^{u v}=G^{v u} =2 \, .
\ee
Equivalently, the dual (vectorial) basis 
$\left\{\alpha_{\perp} , \alpha_u, \alpha_v \right\} =\left\{\alpha_{\perp}^p \frac{\partial}{\partial \beta^p} , \alpha_u^p \frac{\partial}{\partial \beta^p}, \alpha_v^p \frac{\partial}{\partial \beta^p} \right\} $ of $\left\{\alpha^{\perp}, \alpha^u, \alpha^v \right\}$
is equal to $\left\{\alpha_{\perp} , \alpha_u, \alpha_v \right\} =\left\{ \frac12 \alpha^{\sharp}, \frac12 v^{\sharp}, \frac12 u^{\sharp} \right\}$, and the nonzero basis components of the covariant metric $G_{\widehat a  \, \widehat b}$ are 
\be
G_{\perp \perp} =  G_{u v}=G_{v u} =\frac12 \, .
\ee
When considering the one-wall approximation, the potential terms $\widehat V_A(\beta,\Phi) $ entering the 
susy constraints \eqref{SA}, \eqref{susy}, are easily seen to depend on the $\beta$'s only through the single
combination $\alpha(\beta)$. This immediately implies that the two wall-parallel components $\pi_u= - i \frac{\partial}{\partial u}$, 
$\pi_v= - i \frac{\partial}{\partial v}$ of the momentum are conserved. Actually, it is better to consider the parallel components
of the {\it shifted} momentum operator, i.e. the differentiation operator acting on the rescaled wave function $F(\beta)^{-1}  |\Psi \rangle$,
 i.e. $\pi'_a = \pi_a + i \partial{ \ln F}/\partial{ \beta^a}$.  When considered possibly near the wall $\alpha_{23}$, but far from the two other
 walls, the scale factor $F(\beta)$, \eqref{F}, reads (as a function of $\alpha\equiv\alpha_{23}, u, v$)
 \be
 F(\alpha, u, v) \approx e^{-\frac 12 u+\frac 38 v} (2  | \sinh \alpha |)^{-1/8} \,.
 \ee
 Hence, the part of $\ln F(\alpha, u, v)$ that depends on $u$ and $v$ is ${-\frac 12 u+\frac 38 v}$, and shifts the 
 conserved parallel momenta according to: $\pi'_u = \pi_u - \frac12 i $, $\pi'_v = \pi_v +  \frac38 i $. In keeping with the
 type of wavelike solutions (bouncing between potential walls) we are interested in, we shall henceforth consider wave packets
 having real values of the shifted conserved momenta $\pi'_u$,  $\pi'_v$ (and therefore complex values of $\pi_u$, and $\pi_v$).
 
 Putting together the ingredients we just discussed (adapted coordinates, adapted basis, conserved shifted parallel momenta), we finally look for solutions of the susy constraints \eqref{SA}, in the
 one-wall approximation, of the form
 \be \label{onewallstate1}
 |\Psi \rangle_{{\bf 6}, N_F=2}= e^{(i\pi'_u-\frac 12) u+ (i \pi'_v +\frac 38) v}  |{\mathcal F}(\alpha)\rangle_{{\bf 6}, N_F=2}\,,
 \ee
 where
 \be \label{onewallstate2}
 |{\mathcal F}(\alpha)\rangle_{{\bf 6}, N_F=2} = K_{\widehat a \, \widehat b}(\alpha) \tilde b_+^{( \widehat a}\tilde b_-^{\widehat b)} \, \vert 0 \rangle_-  \,.
 \ee
 Inserting this expression in the (one-wall-approximated) susy constraints \eqref{SA}, \eqref{susy} leads to constraints on 
 $ |{\mathcal F}(\alpha)\rangle_{{\bf 6}, N_F=2}$ of the form (with $\pi_u = \pi'_u + \frac{i}{2}$, $\pi_v=\pi'_v - \frac{3 i}{8}$)
\begin{equation} \label{SAEqs2}
\frac i2 \, \widehat \Phi_A^{\perp}  \, \partial_{\alpha} \,  |{\mathcal F}(\alpha)\rangle 
+ \left(- \frac12 \pi_u \widehat \Phi_A^u -\frac12 \pi_v  \widehat \Phi_A^v 
 + \widehat V_A(\alpha,\Phi)  \right) 
 |{\mathcal F}(\alpha)\rangle = 0\quad .
\end{equation}
We recall that the spinor index $A$ takes four values. For each value of $A=1,2,3,4$, Eq. \eqref{SAEqs2} is a Diraclike equation
for the quantum spinor state $|{\mathcal F}(\alpha)\rangle$, with $\Phi_A^{\perp}$ playing the role of a gamma matrix controlling
the evolution with respect to $\alpha$. The anticommutation law \eqref{clifford}
implies 
\be
\widehat\Phi_A^\perp  \widehat\Phi_B^\perp +  \widehat\Phi_B^\perp  \widehat\Phi_A^\perp= \delta_{AB} \, G^{\perp \perp} \, {\rm Id}= 2\, \delta_{AB} \,  \, {\rm Id} \, ,
\ee
so that we see that each matrix  $\widehat \Phi_A^{\perp}$ is invertible (with itself as inverse). 
Multiplying each one of the four equations \eqref{SAEqs2} by $\frac{2}{i} \widehat \Phi_A^{\perp}$ yields
 an overdetermined system of ordinary (matrix) differential equations in $\alpha$ of the form
\begin{equation}\label{delSig}
\partial_{\alpha} \,  |{\mathcal F}(\alpha)\rangle = \widetilde\Sigma_A \,  |{\mathcal F}(\alpha)\rangle \qquad (A=1,\ldots ,4) \qquad .
\end{equation}
The  unknowns of this system are the six components $K_{\widehat a \, \widehat b}(\alpha)$ parametrizing the state  
\eqref{onewallstate1}, \eqref{onewallstate2}. Considering the differences between the equations  \eqref{delSig}, we see that
 the six  components $K_{\widehat a \, \widehat b}(\alpha)$ are subject to the following system of  linear equations
\begin{equation}\label{compaS}
\left(\widetilde\Sigma_1 - \widetilde\Sigma_A \right) |{\mathcal F}\rangle = 0 \qquad (A=2,3,4) \qquad .
\end{equation}
We found that the rank of this linear system\footnote{\label {fnote}Eq. (\ref{delSig}) also leads to six more algebraic constraints, because the operators $\widetilde\Sigma_A$ map the $ |{\mathcal F} \rangle$--components partially outside the subspace to which they belongs. However these extra conditions are found to be consequences of Eqs (\ref{compaS}); similar dependences also occur when the same analysis is performed at level $N_F=3$, as well as in the other one-gravitational-wall approximations, at levels $N_F=2,\ {\rm or\ } 3$.} is equal to 2. 
In other words, the six components $K_{\widehat a \, \widehat b}$  can be expressed as linear combinations of two of them, chosen for instance as  $K_{\perp\perp}$ and $K_{\perp v}$. It is then useful to parametrize the $\alpha$ dependence of $K_{\perp\perp}$ and $K_{\perp v}$
in terms of two other functions $F(\alpha)$, $ G(\alpha)$, as follows (we henceforth work on the half-line $\alpha > 0$)
\begin{align}
K_{\perp\perp}(\alpha) = C_F\sinh^{3/8}(\alpha)\,F(\alpha) \,,\nonumber \\
K_{\perp v}(\alpha) = C_G  \sinh^{3/8}(\alpha)\,G(\alpha) .  
\end{align}
By appropriately choosing the ratio $C_F/C_G$ between the proportionality constants, we obtain a linear system for
the two functions $ F $ and $ G $ which reads
\begin{equation}\label{dfg}
\partial_{\alpha} \, F = G \qquad ,
\end{equation}
\begin{equation}\label{dgf}
\left[\partial_{\alpha} + \coth (\alpha) \right] \, G = -( \frac14 + \pi'^2_{\perp}) \, F  \qquad .
\end{equation}
Here, $\pi'^2_{\perp}$ denotes the function of $\pi'_u, \pi'_v$ defined by  the far-wall $N_F=2$ mass-shell constraint
\begin{equation} \label{defpiperp}
G^{\widehat a \widehat b} \pi'_{\widehat a} \pi'_{\widehat b} = 2 \, \pi^{\prime\,2}_\perp +4\,  \pi'_u\, \pi'_v  = \frac38 \qquad .
\end{equation}
The general solution of the differential system \eqref{dfg}, \eqref{dgf}, contains two arbitrary constants, say $C_P$ and $C_Q$.
The solution parametrized by $C_Q$ involves $Q$-type Legendre functions, and is singular (in a non square-integrable way) on
the considered symmetry wall $\alpha=0$. [E.g., $K_{\perp v}(\alpha) \propto  C_Q \sinh^{3/8}(\alpha)\,
\,Q^{1}_{\nu}[\cosh(\alpha)]$  blows up like $O[\alpha^{-5/8}]$ when $\alpha \to 0$.]
In keeping with the general aim of our work, we shall only consider here the solution parametrized
by $C_P$ which involves $P$-type Legendre functions, which vanish on the symmetry wall. 

Here we use Legendre functions   defined on the complex plane cut between $z=-1$ and $z=1$ by analytically continuing the expression
\be\label{defP}
P_{\nu}^{\mu} [z] = \frac1{\Gamma (1-\mu)} \left( \frac{z+1}{z-1} \right)^{\mu/2} \, _2F_1 \left[ -\nu , 1+\nu ; 1-\mu ; \frac{1-z}2 \right] \qquad .
\ee
$F$ and $G$ involve Legendre functions of order $\mu= 0$ or $\mu=1$, and degree $\nu= -\frac 12+i\,\pi'_\perp$.
Here, we conventionally define $\pi'_\perp$ as the positive solution of the far-wall mass-shell condition \eqref{defpiperp}.
More precisely
\begin{eqnarray}\label{PnuNF2}
F &= &C_P \, P_{-\frac 12+i\,\pi'_\perp}^0 [\cosh (\alpha)] \label{fP0} \,,\\
G &= &C_P \, P_{-\frac 12+i\,\pi'_\perp}^1 [\cosh (\alpha)] \label{gP1} \,. 
\end{eqnarray}
Note that while the definition \eqref{defP} can be used as is when $\mu=0$, the case $\mu=1$ involves a ``regularized"
hypergeometric function (where the vanishing pre-factor $1/\Gamma (1-\mu)$ is needed to regularize the singular coefficients $1/(1-\mu)$
entering the hypergeometric series).

Finally, the general (square-integrable) solution of the $N_F=2$ susy constraints is of the form \eqref{onewallstate1},  \eqref{onewallstate2},
with adapted-basis components $K_{\widehat a \, \widehat b}(\alpha)$ given by (with $\nu= -\frac 12+i\,\pi'_\perp$)
\begin{eqnarray}
\label{FSpepe}
K_{\perp\perp}(\alpha) &= &C_P \, K^0_{\perp\perp} [\pi'_u,\pi'_v] \sinh^{3/8} (\alpha) \, P_{\nu}^0 [\cosh (\alpha)]\qquad, \\
\label{FSpapa}
K_{U V}(\alpha)&= &C_P\, K^0_{U V} [\pi'_u,\pi'_v] \sinh^{3/8} (\alpha) \, P_{\nu}^0 [\cosh (\alpha)]\qquad ,  \\
\label{FSpape}
K_{\perp U}(\alpha) &= &C_P \, K^0_{\perp U} [\pi'_u,\pi'_v]  \sinh^{3/8} (\alpha) \, P_{\nu}^1 [\cosh (\alpha)]\qquad .
\end{eqnarray}
Here, the indices $U,V$ run over the two values $u,v$ parametrizing the parallel components of the wave function, and the
$\pi'_U$-dependent (but $\alpha$-independent) polarization tensors $K^0_{\widehat a \, \widehat b}(\pi'_U)$ are given by
\begin{eqnarray}
\label{Kpepe}
K ^0_{\perp\perp}[\pi'_u,\pi'_v] &= &(\pi'_u+\frac i4)(\pi'_v-\frac i8)  \,,\\
\left\{K^0_{U V}[\pi'_u,\pi'_v]\right\}_{UV=uu,uv,vv} &= &  \left\{-\frac 12\pi_v^{\prime 2}+i\frac 7{16} \pi'_v+i\frac 3{32}\pi'_v-\frac3 {128},\  -\pi'_u\pi'_v+\frac i 2\left(\pi'_u-\pi'_v)\right)-\frac{11}{32},\right.  \nonumber\\
&& \ \left.\  - \frac 12 \pi^{\prime 2}_u+i\left(\frac 1{4}\pi'_v-\frac 78\pi'_u\right)+\frac {13}{32} \right\} \,, 
\label{Kpapa}\\
\label{Kpape}
\left\{K^0_{\perp U}[\pi'_u,\pi'_v]\right\}_{U=u,v}&= &\left\{ (i\,\pi'_v+\frac 18),\   (i\,\pi'_u- \frac 34)\right\} \,.
\end{eqnarray}
We have checked that the values of the various $\pi'_U$-dependent coefficients $K^0_{\widehat a \, \widehat b}(\pi'_U)$ are
in agreement with the general, far-wall plane-wave solution \eqref{farwall}. To perform this check, and to finally obtain the scattering laws \eqref{reflection1}, 
 \eqref{reflection2},  \eqref{reflection3}, we will need to use  the far-wall ($\alpha \to + \infty$) asymptotic expression of the 
 Legendre functions, namely:
\begin{equation}\label{PmuAs}
P^\mu_\nu[\cosh(\alpha)]{\approx}\frac 1{\sqrt{\pi}}\left(\frac{\Gamma(\frac 12+\nu)}{\Gamma(1-\mu+\nu)}e^{\nu\, \alpha}+\frac{\Gamma(-\frac 12-\nu)}{\Gamma(-\mu-\nu)}e^{-(\nu+1) \,\alpha}\right)\qquad ,\ \alpha  \rightarrow +\infty \,.
\end{equation}
 
 \subsection{Reflection laws on the symmetry wall  $\alpha_{23}(\beta) = \beta^3- \beta^2$.}
 
Let us now extract from the explicit structure of the one-wall solution \eqref{FSpepe}, \eqref{FSpapa}, \eqref{FSpape} the reflection
laws \eqref{reflection1},  \eqref{reflection2},  \eqref{reflection3}. In the following, we shall conventionally assume that the wavepackets
we are considering are ``future-directed" in the sense that the (shifted, far-wall) contravariant momentum vector $\pi^{\prime \sharp}$ is directed towards
increasing values of the timelike variable $\beta^0= \beta^1 + \beta^2 + \beta^3$. [Physically, as $\beta^0 = - \ln (abc)$,
this means that we are considering a contracting universe, going towards a Big-Crunch-like singularity where the volume $a b c \to 0$.]
With this convention, and given the fact that the wavepacket evolves in the half-space $\alpha=\alpha_{23} >0$,  the ingoing piece of the asymptotic solution is characterized by having a complex phase factor $\propto e^{-i\,\pi'_\perp\,\alpha }$, while its reflected piece should
have a phase factor $\propto e^{+i\,\pi'_\perp\,\alpha }$. Here, as above, $\pi'_\perp$ is defined as being the positive root
of the mass-shell condition \eqref{defpiperp}. 

In the case of the $\alpha_{23}$ symmetry wall that interest us here, we should insert $\nu=- \frac12+i\,\pi'_\perp$ in the asymptotic
expression \eqref{PmuAs}. This yields
\begin{eqnarray}
P^0_{-\frac 12+i\,\pi'_\perp}&\simeq &\frac{e^{-\frac 12 \alpha}}{\sqrt{\pi} } \left( \frac{\Gamma(-i\,\pi'_\perp)}{\Gamma(\frac12-i\,\pi'_\perp)}\, e^{-i\,\pi'_\perp\,\alpha }+   \frac{\Gamma(+i\,\pi'_\perp)}{\Gamma(\frac12+i\,\pi'_\perp)} e^{+i\,\pi'_\perp\,\alpha }\right)\,, \\
P^1_{-\frac 12+i\,\pi'_\perp}&\simeq & \frac{e^{-\frac 12 \alpha}}{\sqrt{\pi} } \left( \frac{\Gamma(-i\,\pi'_\perp)}{\Gamma(-\frac12-i\,\pi'_\perp)}\, e^{-i\,\pi'_\perp\,\alpha }+   \frac{\Gamma(+i\,\pi'_\perp)}{\Gamma(-\frac12+i\,\pi'_\perp)} e^{+i\,\pi'_\perp\,\alpha }\right) \,.
\end{eqnarray}

Let us first note that the combination of the exponentially decaying prefactor $e^{ -\frac 12\,\alpha }$ 
with the overall factor $\sinh^{3/8} (\alpha)$ in Eqs. \eqref{FSpepe}, \eqref{FSpapa}, \eqref{FSpape}, and with the real exponential
factor linked to the imaginary additions to $\pi'_u, \pi'_v$ in Eq. \eqref{onewallstate1}, reproduces
(in the limit $\alpha \gg 1$) the real exponential factor $ e^{\beta^1+\frac 34\beta^2+\frac 12\beta^3} =e^{ - \frac18 \alpha-\frac 12 u+\frac 38 v}$ present in the general far-wall solution \eqref{farwall}.
Then, the presence of the two complex-conjugated phase factors $e^{\pm i\,\pi'_\perp\,\alpha }$ (in addition to the conserved
phase factors $e^{ i\,(\pi'_u u + \pi'_v v ) }$) shows that the reflection law for the shifted momentum reads
\be
 \pi'^{\rm in}_{\widehat a}=(-\pi'_\perp, \pi'_u,\pi'_v ) \, \to \, \pi'^{\rm out}_{\widehat a}=(+\pi'_\perp, \pi'_u,\pi'_v ) \,.
 \ee
 The rewriting of this adapted-basis reflection law, precisely yields the specular reflection law \eqref{reflection1}.
 
 In order to extract the global reflection phase-factor $e^{i \delta_{\alpha}^{\rm global}}$, Eq. \eqref{reflection2}, connecting the incident far-wall amplitude to the reflected one, one needs to compare both the incident and the reflected pieces of the solution  Eqs. \eqref{FSpepe}, \eqref{FSpapa}, \eqref{FSpape} to the generic far-wall solution \eqref{farwall}. When doing so, one can first factor out the amplitude
 of, say, the incident $P^0_{-\frac 12+i\,\pi'_\perp}$-type modes (in $K_{\perp\perp}$ and $K_{UV}$). This yields a
  $\pi'_\perp$-dependent factor in the 
 corresponding incident $P^1_{-\frac 12+i\,\pi'_\perp}$-type modes (in $K_{\perp U}$) given by
 \be \label{pi'factor}
 \frac{ \Gamma(\frac12-i\,\pi'_\perp) }{ \Gamma(-\frac12-i\,\pi'_\perp) }= -\frac12-i\,\pi'_\perp \, ,
 \ee
 where we used the basic identity $\Gamma(z+1)= z \, \Gamma(z)$. Combining this additional $\pi'_\perp$-linear factor \eqref{pi'factor}
 in $K_{\perp U}$ with the $\pi'_U$-linear factors displayed in Eq. \eqref{Kpape}, we found that all the $K_{\widehat a \widehat b}$
 incident amplitudes of Eqs. \eqref{FSpepe}, \eqref{FSpapa}, \eqref{FSpape} nicely agree with the $\pi'_a$-quadratic dependence
 of the generic far-wall amplitude  \eqref{farwall} derived in our previous work \cite{Damour:2014cba}, and recalled
 in Appendix A. [The same check holds
 for the reflected amplitude.]

 As additional result of this asymptotic analysis, one gets the relation between
 the overall scalar amplitude $C_{N_F=2}$ of a far-wall wave packet and the overall coefficient $C_P$ parametrizing
 the amplitude of the $P$-type solution, namely
 \be
 C_{N_F=2}^{\pm} = - \frac{C_P }{ \sqrt{\pi} 2^{\frac{11}{8}} }\frac{\Gamma(\pm i\,\pi'_\perp)}{\Gamma(\frac12 \pm i\,\pi'_\perp)} \quad .
 \ee
 where the upper sign on $C_{N_F=2}$ refers to the outgoing wave (having $ \alpha_{23}\cdot \pi' >0$), while the lower sign
 refers to the ingoing wave. Taking the ratio between $C_{N_F=2}^{+}$ and $C_{N_F=2}^{-}$ yields the global phase factor
 \be \label{globphase23}
  e^{i \delta_{\alpha_{23}}^{\rm global}}=\frac{C_{N_F=2}^{+}}{C_{N_F=2}^{-}}= \frac{\Gamma[+i\pi'_\perp]\,\Gamma[\frac 12 -i\,\pi'_\perp]}{\Gamma[-i\pi'_\perp]\,\Gamma[\frac 12 +i\,\pi'_\perp]} \quad .
 \ee
 In the small wavelength (WKB) limit  (large values for the components $\pi'_a$), this yields, using
 \be
 \frac{\Gamma(z+a) }{\Gamma(z+b)} \approx z^{a-b} \, {\rm as} \,   z \to \infty \quad ,
 \ee
 \be \label{globphasewkb}
 e^{i \delta_{\alpha_{23}}^{\rm global, WKB}} \approx e^{- i \frac{\pi}{2}} \quad .
 \ee
 The latter asymptotic value of the global phase is also easily obtained by considering  the $K _{\perp\perp}$ component of the $N_F=2$ solution
 (which is given, for large values of $\pi'_a$, by $K _{\perp\perp} \approx C_{N_F=2} \pi'_{\perp} \pi'_{\perp}$ where the factor 
 ${\pi'_{\perp}}^2$ does not change sign upon reflection).
 
 Finally, let us extract from our results above the reflection operator (in Hilbert space) $\mathcal{ R}_\alpha^{{\bf 6}, N_F=2}$
 mapping the incident state $|\Psi \rangle^{\rm in}_{{\bf 6}, N_F=2}$ to the reflected one. We can compute this operator by
 relating the various basis spinor states $\tilde b_+^{( \widehat a}\tilde b_-^{\widehat b)} \, \vert 0 \rangle_- $ making up the
 one-wall solution \eqref{onewallstate2} to eigenstates of various operators defined in terms of the basic gravitino operators 
 $\widehat \Phi^a_A$. Let us recall that our previous work had emphasized that the building blocks of the susy Hamiltonian operator were
 some operators quadratic in the $\widehat \Phi^a_A$'s that generated a representation of the compact subalgebra $K[AE_3]$ 
 of $AE_3$. There were two types of such operators: the three spin operators $\widehat S_{12}$, $\widehat S_{23}$, $\widehat S_{31}$,
 associated with symmetry walls, and three operators $\widehat J_{11}$, $\widehat J_{22}$, $\widehat J_{33}$, associated
 with the three dominant gravitational walls $\alpha_{11}= 2 \beta^1$, $\alpha_{22}= 2 \beta^2$,  $\alpha_{33}= 2 \beta^3$.
[See Eq. (8.10) in Ref. \cite{Damour:2014cba}.] 
Here, we are considering the reflection by the symmetry wall $\alpha_{23}$, so that one might expect that the corresponding
reflection operator $\mathcal{ R}_\alpha^{{\bf 6}, N_F=2}$ might be directly related to the corresponding spin operator $\widehat S_{23}$.
There is, however, a subtlety. Indeed, while the considered dynamical states $|\Psi \rangle_{{\bf 6}, N_F=2}$ live in a 6-dimensional
subspace ${\mathbb H}_{(1,1)_S}$ of the 15-dimensional $N_F=2$ level (so that $\mathcal{ R}_\alpha^{{\bf 6}, N_F=2}$ is an endomorphism of ${\mathbb H}_{(1,1)_S}$), the spin operator $\widehat S_{23}$ happens not to leave invariant ${\mathbb H}_{(1,1)_S}$, but to map
it to other sectors within the 15-dimensional $N_F=2$ state space. However, if one considers, instead of $\widehat S_{23}$, its
square, namely $\widehat S_{23}^2$, one checks that the latter operator leaves invariant (and thereby defines an endomorphism of)
  ${\mathbb H}_{(1,1)_S}$. [We recall
that it is indeed the squared spin operator which enters each symmetry wall $\alpha_{ab}$ in the Hamiltonian operator, as per $\sim (\widehat S_{ab}^2-Id)/(4\,\sinh^2(\alpha_{ab}))$.] In addition, we have shown that the basis of spinor states $\tilde b_+^{( \widehat a}\tilde b_-^{\widehat b)} \, \vert 0 \rangle_- $ entering \eqref{onewallstate2}, and which were crucial for finding and simplifying the solution
of the susy constraints happen to be eigenstates of $\widehat S_{23}^2$. More precisely, we have shown that  four of our basis states
are eigenstates of $\widehat S_{23}^2$ with zero eigenvalues,
\be
\widehat S_{23}^2 \, \tilde b_+^{ \perp}\tilde b_-^{\perp} \, \vert 0 \rangle_-  = 0 \,;    \,
 \widehat S_{23}^2 \, \tilde b_+^{ (U}\tilde b_-^{V)} \, \vert 0 \rangle_-  = 0 \,,
\ee
while the other two basis states are eigenstates of $\widehat S_{23}^2$ with  eigenvalue equal to 4:
\be
\widehat S_{23}^2 \, \tilde b_+^{ (\perp}\tilde b_-^{U)} \, \vert 0 \rangle_-  = 4  \, \tilde b_+^{ (\perp}\tilde b_-^{U)} \, \vert 0 \rangle_-  \, .
\ee
We then note that these eigenvalues of $\widehat S_{23}^2$ are correlated to the Legendre order $\mu$ of the corresponding wavefunction
$K_{\widehat a \, \widehat b}(\alpha)$ by the simple rule
\be \label{muS23}
\mu = \frac12 |\widehat S_{23}|_{{\bf 6}, N_F=2} \,,
\ee
where we introduced the operator  $|\widehat S_{23}|_{{\bf 6}, N_F=2}$ defined as being the (unique) positive square 
root\footnote{By definition, we require this square root to have the same eigenstates as ${\widehat S_{23}^2}{}_{{\bf 6}, N_F=2}$.} of 
$\widehat S_{23}^2$, considered as an endomorphism of ${\mathbb H}_{(1,1)_S}$.

When comparing the phases of the incident and reflected pieces in the one-wall solution above, one easily sees that they
only differ by a phase factor, and that the latter phase factor, say $e^{i \delta_\mu}$ only depends on the value of the Legendre order $\mu$,
and can be written as 
\begin{equation} \label{deltamu}
e^{i\,\delta_\mu}=\frac{\Gamma(\frac12 - \mu -i\,\pi'_\perp)\,\Gamma( i\,\pi'_\perp)}{\Gamma(\frac12 - \mu +i\,\pi'_\perp)\,\Gamma(-i\,\pi'_\perp)} \quad .
\end{equation}
In view of the
strict correlation \eqref{muS23}, we conclude that the reflection operator $\mathcal{ R}_{\alpha_{23}}^{{\bf 6}, N_F=2}$
is an operatorial function of $|\widehat S_{23}|_{{\bf 6}, N_F=2}$, which is given by
\be \label{Ra23NF2}
\mathcal{ R}_{\alpha_{23}}^{{\bf 6}, N_F=2} = \frac{\Gamma[+i\pi'_\perp]\,\Gamma[\frac 12 -i\,\pi'_\perp- \frac12|\widehat S_{23}|_{{\bf 6}, N_F=2}]}{\Gamma[-i\pi'_\perp]\,\Gamma[\frac 12 +i\,\pi'_\perp- \frac12 |\widehat S_{23}|_{{\bf 6}, N_F=2}]}\qquad .
\ee
In the small wavelength (or WKB) limit ($\pi'_\perp \gg 1$), we have
\be
\left[ e^{i\,\delta_\mu} \right]_{\rm WKB} = e^{i \pi (\mu-\frac12)} \,,
\ee
so that the reflection operator depends only on the spin operator, namely
\be \label{Ra23NF2WKB}
\mathcal{ R}_{\alpha_{23}}^{{\bf 6}, N_F=2, WKB} = e^{ +  \frac{ i\pi}{2} (|\widehat S_{23}|_{{\bf 6}, N_F=2}-1)}= e^{i \delta_{\alpha_{23}}^{\rm global}} \, e^{ + \frac{ i\pi}{2} |\widehat S_{23}|_{{\bf 6}, N_F=2} } \,.
\ee
In the second form, we have factored out the (WKB limit of the) global phase factor \eqref{globphasewkb} (which corresponds
to the $\widehat S_{23}^2=0$ eigenvalues). Note that this result can also be written as
\be
\mathcal{ R}_{\alpha_{23}}^{{\bf 6}, N_F=2, WKB} = e^{-\frac{ i\pi}{2}} \, e^{ - \frac{ i\pi}{2} |\widehat S_{23}|_{{\bf 6}, N_F=2} } \,,
\ee
because the eigenvalues of $ |\widehat S_{23}|_{{\bf 6}, N_F=2}$ are 0 and 2.

\subsection{Scattering and reflection laws on the symmetry wall  $\alpha_{12}(\beta) = \beta^2- \beta^1$.}

We shall be briefer in our discussion of the scattering of a $N_F=2$ wave packet \eqref{kpq615} on the other symmetry
wall of our canonical chamber, i.e. the wall form $\alpha_{12}(\beta) = \beta^2- \beta^1$. Though 
there are some differences in intermediate expressions
(because of the dissymetric role of the two symmetry walls bounding one given  billiard chamber)
the final results are obtained by applying the cyclic permutation $ (231) \to (123)$ to the previous final results concerning
the scattering on the $\alpha_{23}(\beta) = \beta^3- \beta^2$ wall.

Again, the crucial tool is to work within a basis of one forms adapted to the considered wall. The previous basis \eqref{basisNF2a23},
with \eqref{a23}, \eqref{uv} is now replaced by
\be \label{basisNF2a12}
\left\{\tilde \alpha^{\perp}, \tilde \alpha^u, \tilde \alpha^v \right\} \equiv \left\{\tilde \alpha(\beta), \tilde u(\beta), \tilde v(\beta)\right\} \quad ,
\ee
with
\begin{equation} \label{tildedbasis}
\tilde\alpha(\beta) =\beta^2-\beta^1\quad ,\quad 
\tilde u(\beta)\equiv-(2\,\beta^3 + \frac 12\,\beta^1+\frac 12\,\beta^2)\quad ,\quad
\tilde v(\beta)\equiv(\beta^1+\beta^2)\quad .
\end{equation}
The metric components in this adapted basis are the same as the previous ones, Eq. \eqref{Gup23}, so that the far-wall mass-shell
condition reads as before, namely
\begin{equation} \label{defpiperptilde}
G^{\widehat a \widehat b} \pi'_{\widehat a} \pi'_{\widehat b} = 2 \, \pi^{\prime\,2}_\perp +4\,  \pi'_{\tilde u}\, \pi'_{\tilde v}  = \frac38 \qquad .
\end{equation}
\be
\ee
The state reflecting on the $\alpha_{12}$ wall is looked for in the form
\be \label{onewallstate1a12}
 |\Psi \rangle_{{\bf 6}, N_F=2}= e^{(i\pi'_{\tilde u}-\frac 14) \tilde u+ (i \pi'_{\tilde v} +\frac 34) \tilde v}  |\tilde {\mathcal F}(\alpha)\rangle_{{\bf 6}, N_F=2} \,,
 \ee
 where the real-exponential contributions are slightly modified (because of the non cylic invariance of the original scale factor $F(\beta)$, Eq. \eqref{F}), and where
 \be \label{onewallstate2a12}
 |\tilde {\mathcal F}(\alpha)\rangle_{{\bf 6}, N_F=2} = K_{\widehat a \, \widehat b}(\alpha) \tilde b_+^{( \widehat a}\tilde b_-^{\widehat b)} \, \vert 0 \rangle_-  \,.
 \ee
Here it is now understood that the basis indices $ \widehat a=(\perp, u, v)$ must be replaced by their tilded avatars,
corresponding to the new basis \eqref{basisNF2a12}, \eqref{tildedbasis}.

As above, we find   $Q^{\mu}_{\nu}$-type and  $P^{\mu}_{\nu}$-type Legendre solutions, with the order $\mu$ related to  
${\widehat S_{12}}^2$ via
\be
\mu = \frac12 |\widehat S_{12}|_{{\bf 6}, N_F=2}\quad ,
\ee
so that $\mu=0$ or $1$. The degree $\nu$ is again given by $\nu= -\frac12 + i \,\pi'_\perp$.  The $Q$-type solutions are singular
and we discard them.  On the other hand, the $P$-type solutions are regular and are expressed by formulas similar to
Eqs. \eqref{FSpepe}, \eqref{FSpapa}, \eqref{FSpape}, when using projections on the tilded basis \eqref{basisNF2a12}.

The final reflection laws are the same, {\it mutatis mutandis}, as before. Namely, the standard specular reflection law \eqref{reflection1}
(on the new wall $\alpha_{12}$), and (defining as before $\pi'_\perp$ as the positive root of  the mass-shell condition \eqref{defpiperptilde})
\be \label{globphase12}
  e^{i \delta_{\alpha_{12}}^{\rm global}}= \frac{\Gamma[+i\pi'_\perp]\,\Gamma[\frac 12 -i\,\pi'_\perp]}{\Gamma[-i\pi'_\perp]\,\Gamma[\frac 12 +i\,\pi'_\perp]} \approx e^{-\frac{ i\pi}{2}}
 \ee
 (where the last approximation corresponds to the WKB limit)
 and the $(23) \to (12)$ version of the reflection operator \eqref{Ra23NF2}, which yields, in the WKB limit
 \be \label{Ra12NF2WKB}
\mathcal{ R}_{\alpha_{12}}^{{\bf 6}, N_F=2, WKB} = e^{-\frac{ i\pi}{2}}
e^{ \pm  \frac{ i\pi}{2} |\widehat S_{12}|_{{\bf 6}, N_F=2}} \,.
\ee
 As before, we can indifferently choose here the $\pm$ sign because the eigenvalues of $ |\widehat S_{12}|_{{\bf 6}, N_F=2}$ are, as
 before, $0$ and $2$.

\subsection{Scattering on the gravitational wall $\alpha_{11}(\beta) = 2 \beta^1$.}

We shall also be brief in discussing the scattering of a $N_F=2$ wave packet \eqref{kpq615} on a gravitational wall. 
Gravitational walls correspond to terms in the Hamiltonian that are proportional to $e^{- \alpha_{11}} =  e^{- 2 \beta^1} $,
$e^{- \alpha_{22}} =  e^{- 2 \beta^2}$, or $e^{- \alpha_{33}} =  e^{- 2 \beta^3}$. 
The main differences between a gravitational wall  and  a symmetry wall  are that: (i) a gravitational wall is softer than a symmetry wall 
in that it does not become singular on the corresponding wall hyperplane $\alpha_{aa}=0$;
 and (ii) the operator $\widehat J_{11}$ coupled (in the
Hamiltonian) to the wall factor $e^{- \alpha_{11}}$ is  quadratic in the gravitino operators $\widehat \Phi^a_A$, while we
had quartic-in-fermions operators, such as $\widehat S_{23}^2$, for symmetry walls (see Eq. (8.11) in Ref. \cite{Damour:2014cba}).  
Similarly to the (sharper) symmetry wall case, we shall impose  the boundary condition that the wave function exponentially decreases 
as one penetrates within the considered gravitational wall (i.e. when, say, $\alpha_{11}(\beta) = 2 \beta^1$ becomes negative).

As in the symmetry-wall case, we shall solve the susy constraints in the one-wall approximation. It is again very useful to introduce
an adapted basis of one-forms, namely $\,_g\alpha^{\widehat a} =(\,_g\alpha^{\perp}, \,_g\alpha^u,\,_g\alpha^v)$ with
\be\label{gbasis}
\,_g\alpha^{\perp}(\beta)=2 \beta^1\,;\, \,_g\alpha^u(\beta)=\,_gu=\beta^1+\beta^3 \,;\,_g\alpha^v(\beta)= \,_gv=\beta^1+\beta^2\qquad 
\ee
[Below, we simplify the notation by deleting the pre-subscript $g$.]
Again, we have chosen a direction normal to the considered wall, and two null directions parallel to the wall.
The normalization of this co-frame is now slightly different from before, with
\be
G^{\perp \perp} = 2 \, ; \,  G^{u v}=G^{v u} =-1 \, ,
\ee
so that the far-wall mass-shell condition reads
\be
2\pi^{\prime 2}_\perp - 2 \pi'_u\,\pi'_v=\frac 38   \, .
\ee
In the following, we shall define $\pi'_{\perp}$ as the positive root of the latter mass-shell condition, i.e. $\pi'_\perp=\sqrt{\pi'_u\,\pi'_v - \frac12 \mu^2}$, where $\mu^2= - \frac38$ is the squared mass at level $N_F=2$.
The dual (vectorial) basis is equal to $\left\{\alpha_{\perp} , \alpha_u, \alpha_v \right\} =\left\{ \frac12 \alpha^{\sharp}, - v^{\sharp}, - u^{\sharp} \right\}$. 

Let us introduce the shorthand notation (here generalized, in anticipation of the corresponding $N_F=3$ discussion, to a mass-shell
condition involving a different squared mass $\mu^2$)
\begin{equation} \label{Uj}
{\mathcal U}_{\mu^2} (\beta ; j) \equiv e^{\left( \frac34 \beta^2 + \frac12 \beta^3 \right)} \,e^{i \pi'_u\, (\beta^1+\beta^3 )+ i\pi'_v\,(\beta^1+\beta^2 ) }  \, e^{2\,\beta^1} \,\,
W_{-  \frac12 j,\,i\,\pi'_\perp}[ e^{-2\beta^1}] \,, 
\end{equation}
where $W_{\kappa, \mu}(z)$ denotes the standard Whittaker function.

The solution of the susy constraints near a gravitational wall (and decaying under the wall) is then of the usual form (see \eqref{kpq615})
\begin{equation}\label{kpq615basis}
|\Psi \rangle_{{\bf 6}, N_F=2}= k_{\widehat a \widehat b}(\beta) \tilde b_+^{(\widehat a} \tilde b_-^{\widehat b)} \, \vert 0 \rangle_-   \, ,
\end{equation}
where the frame indices now refer to the gravitational basis \eqref{gbasis}, and where the components of the state are given by

\begin{eqnarray}
k_{uu}(\beta) & = &  C^J \,{\left(\frac 54 +i(\pi'_u+2\,\pi'_v)-\pi_v^{\prime 2}\right)}  \,{\mathcal U}_{-\frac38}[\beta,- \frac32] \,,\\
k_{uv}(\beta) & = &C^J\, \frac 14 {(1+2\,i\,\pi'_u)(1+2\,i\,\pi'_v)}  \,{\mathcal U}_{-\frac38}[\beta,-\frac32] \,,\\
k_{vv}(\beta) & = &C^J\,\frac 14  {(1+2\,i\,\pi'_u)^2}  \,{\mathcal U}_{-\frac38}[\beta,- \frac32] \,,\\
k_{\perp u}(\beta) & = &-C^J\,\frac 14  {(3+2\,i\,\pi_v')(1+4\,\pi_u'\,\pi_v')}  \, {\mathcal U}_{-\frac38}[\beta,\frac12] \, , \\
k_{\perp v}(\beta) & = &-C^J\,\frac 14{(1+2\,i\,\pi_u')(1+4\,\pi_u'\,\pi_v')} \, {\mathcal U}_{-\frac38}[\beta,\frac12] \, ,\\
k_{\perp\perp}(\beta) & = &-C^J\,  {(1+4\,\pi_u'\,\pi_v' )} \left(\frac 12{\mathcal U}_{-\frac38}[\beta,- \frac32]-\left(\frac 12e^{-2\,\beta^1}+\frac 14\right){\mathcal U}_{-\frac38}[\beta,\frac12]\right) \, . 
\end{eqnarray}
The last component  can be rewritten as
\be
k_{\perp\perp}(\beta) =  {C^J\,\frac18(1+4\,\pi_u'\,\pi_v' )(3+4\, \pi_u'\, \pi_v')} \, {\mathcal U}_{-\frac38}[\beta,\frac52] \, .
\ee
The latter form displays the role of the value $j = \frac52$ in the second argument of the function ${\mathcal U}_{\mu^2} (\beta ; j)$
 describing the behavior of the basis state
$\tilde b_+^{\perp}  \tilde b_-^{\perp} \vert 0 \rangle_- $ (in correspondence with the fact that the latter state is
 an eigenstate of the operator $\widehat J_{11}$ with the eigenvalue
$j = \frac52$).

 The behaviour of the Whittaker function near the origin $e^{- 2 \beta^1} \to 0$ yields the far-wall limit of the wave function:
 \begin{eqnarray}   \label{asympU}
   {\mathcal U}_{\mu^2}[\beta,j]&\underset{\beta^1\rightarrow\infty} \sim& e^{(\beta^1+\frac 34\beta^2+\frac 12\beta^3)}e^{i\left(\pi'_v(\beta^1+\beta^2)+\pi'_u(\beta^1+\beta^3)\right)}\nonumber\\
   & & \times \left(\frac{\Gamma\left[-i\, 2\,\pi'_\perp\right]}{\Gamma\left[\frac{(1+j)}2-i\,\pi'_\perp\right]}e^{-i\,2\,\pi'_\perp\,\beta^1} +\frac{\Gamma\left[i\, 2\,\pi'_\perp\right]}{\Gamma\left[\frac{(1+j)}2+i\,\pi'_\perp\right]}e^{i\,2\,\pi'_\perp\,\beta^1} 
\right) \,.
\end{eqnarray}
We have checked that the $\pi'_\perp$-dependence of the successive ratios between the incident and reflected amplitudes exhibited
in \eqref{asympU}, which follow from the Euler-gamma function identity
\begin{equation}\Gamma\left[\frac{(1+j+2)}2-i\,\pi'_\perp\right] =
 \left(\frac{(1+j)}2-i\,\pi'_\perp\right)\,\Gamma\left[\frac{(1+j)}2-i\,\pi'_\perp\right] \,,
 \end{equation} 
agree with the general far-wall solution \eqref{farwall}, obtained in Ref. \cite{Damour:2014cba}.

From Eq. \eqref{asympU}, we also immediately get the phase shifts, for each component of the wave function, between the incident ($\propto e^{-i\,2\,\pi'_\perp\,\beta^1}$) and reflected ($\propto e^{+i\,2\,\pi'_\perp\,\beta^1}$) amplitudes, upon scattering on the $\alpha_{11}= 2 \beta^1$
 gravitational wall:
\begin{equation} \label{deltaj}
e^{i\delta_{\alpha_{11}}(j ; \pi'_\perp)} = \frac{\Gamma \left[ \frac{1+j}2 - i \,\pi'_\perp \right] \Gamma \left[ i\,2\,\pi'_\perp \right]}{\Gamma \left[ \frac{1+j}2 + i \,\pi'_\perp \right] \Gamma \left[ -i\,2\,\pi'_\perp \right]} \quad .
\end{equation}

In the cases of the reflections upon symmetry walls discussed above, the global phase factor, entering Eq. \eqref{reflection2}, could be read
off from the reflection behavior of the perpendicular-perpendicular amplitude, $k_{\perp\perp}(\beta)$. The reason for this fact was that, in those cases,
 the perpendicular-perpendicular projection of the farwall amplitude \eqref{farwall}, i.e. the quantity $K_{\perp \perp}(\pi'_a)$, happened
 to be independent of the sign of the (corresponding) perpendicular component $\pi'_{\perp}$ of $\pi'_a$ (which is the only adapted-basis 
 component of $\pi'_a$ which changes upon reflection). [The first contribution $\propto {\pi'}_{\perp}^2$ in $K_{\perp \perp}(\pi'_a)$ 
 is always invariant under the sign flip $\pi'_{\perp} \to - \pi'_{\perp}$, but we are talking also here about the second contribution 
 $\propto L_{\perp \perp}^k \pi'_k$, which is linear in $\pi'_{\perp}$ and could, a priori, change under reflection. It does not in the case of
 symmetry-wall reflections because $ L_{\perp \perp}^{\perp}$  happens to vanish.] By contrast, in the case of reflection upon the gravitational wall $\alpha_{11}$,
 we found that the corresponding coefficient $ L_{\perp \perp}^{\perp}$ does not vanish, so that the (linear in $\pi'_a$) contribution $\propto L_{\perp \perp}^k \pi'_k$ changes upon reflection. On the other hand, we found that all the parallel-parallel coefficients $ L_{U V}^{\perp}$
 measuring the dependence on $ \pm \pi'_{\perp}$ vanish in the case of the gravitational wall $\alpha_{11}$. As a consequence, in that case,
 the global phase can be read off  from the reflection behavior of the parallel-parallel amplitudes, $k_{U V}(\beta)$. The latter amplitudes
 correspond to the eigenvalue $j=- \frac32$, so that
\be 
e^{i\delta_{\alpha_{11}}^{\rm global}} = \frac{\Gamma \left[ -\frac14 - i \,\pi'_\perp \right] \Gamma \left[ i\,2\,\pi'_\perp \right]}{\Gamma \left[ -\frac14 + i \,\pi'_\perp \right] \Gamma \left[ -i\,2\,\pi'_\perp \right]} \quad .
\ee
Contrary to the symmetry-wall cases, the dependence of the global reflection phase $\delta_{\alpha_{11}}^{\rm global}$ on $\pi'_\perp$
does not admit a limit when $\pi'_\perp \to + \infty$, rather one has $ \delta_{\alpha_{11}}^{\rm global} \sim \pi'_\perp \ln \pi'_\perp$.
This divergence is easily understood classically: because of the energy conservation law $ {\pi'_\perp}^2 + k\, e^{- 2 \beta^1}={\pi'_\perp}^2_{\rm far-wall}$,
 a relativistic particle impinging on a gravitational wall with incident normal momentum ${\pi'_\perp}_{\rm far-wall}$ will penetrate 
 within the wall up to the  turning point $\pi'_\perp=0$, i.e. up to the energy-dependent position 
 $2 \beta^1_{\rm turning}=- \ln ({\pi'_\perp}^2_{\rm far-wall}/k)$. The shift $ \beta^1_{\rm turning}$ in the effective location
 of the wall then leads to an additional (energy-dependent) phase shift $ \sim - 2 {\pi'_\perp}_{\rm far-wall} \beta^1_{\rm turning}
 \sim 2 {\pi'_\perp}_{\rm far-wall}  \ln {\pi'_\perp}_{\rm far-wall} $. 

Of more importance for our purpose is the dependence of the phase factors \eqref{deltaj} on the second argument $j$ of the
mode function ${\mathcal U}_{\mu^2} (\beta ; j)$, Eq. \eqref{Uj}.  Indeed, we have shown that our adapted
basis was such that each corresponding spinor state $\tilde b_+^{(\widehat a} \tilde b_-^{\widehat b)} \, \vert 0 \rangle_-  $
was an eigenspinor of the operator $\widehat J_{11}$ associated with the $\alpha_{11}$ gravitational wall.
More precisely, the perpendicular-perpendicular, $\perp \perp$, state has eigenvalue $j = \frac52$, the two 
perpendicular-parallel states, $\perp u$, $\perp v$ have eigenvalues $j=\frac12$, and the three parallel-parallel states
$ uu, uv,vv$ have eigenvalues $- \frac32$. Note that these values are precisely the $j$-values entering the corresponding
mode functions \eqref{Uj}. We can therefore re-express the result above by saying that the reflection operator,
in Hilbert space, against the $\alpha_{11}$ gravitational wall is given by the following operatorial expression
\be\label{Ra11NF2}
\mathcal{ R}_{\alpha_{11}}^{{\bf 6}, N_F=2} = \frac{\Gamma \left[ \frac{1+\widehat J_{11}}2 - i \,\pi'_\perp \right] \Gamma \left[ i\,2\,\pi'_\perp \right]}{\Gamma \left[ \frac{1+\widehat J_{11}}2 + i \,\pi'_\perp \right] \Gamma \left[ -i\,2\,\pi'_\perp \right]} \quad .
\ee
In the large momentum (WKB) limit, this yields
\be
\mathcal{ R}_{\alpha_{11}}^{{\bf 6}, N_F=2} \approx e^{i \delta_0(\pi'_\perp)} e^{- i \frac{\pi}{2}}  e^{- i \frac{\pi}{2} \widehat J_{11}} \quad ,
\ee
where we defined $\delta_0(\pi'_\perp) \equiv 2 \pi'_\perp \ln( 4 \pi'_\perp/e)$.

\section{Quantum fermionic billiard at level $N_F = 3$}
\label{SW1020}

The analysis done in the previous section of the various reflection laws at the fermionic level $N_F=2$ can also be 
performed at the fermionic level $N_F=3$. This level corresponds to a 20-dimensional subspace
of the total spinorial state space. Actually, there is a natural decomposition of the $N_F=3$ space into two
10-dimensional subspaces. The latter two subspaces are mapped onto each other via the involution
$b^a_{\pm} \to b^a_{\mp}$,  $\tilde b^a_{\pm} \to \tilde b^a_{\mp}$, between the basic fermionic annihilation and
creation operators. Here, we shall work in only one of these equivalent 10-dimensional subspaces.

As found in our previous work the general structure of the propagating solution of the susy constraints can then
be written as 
\be\label{NF3sol}
|\Psi \rangle_{{\bf 10}, N_F=3} =f(\beta) \, |\eta \rangle +  h_{pq}(\beta) \,   | B^{pq} \rangle ,
\ee
where
\be
 |\eta \rangle = \frac1{3!} \eta_{klm} \, \widetilde b_-^k \, \widetilde b_-^l \, \widetilde b_-^m\,\vert0\rangle_-  \,  ; \,
 | B^{pq} \rangle = \frac12 \sum_{k, l} \eta^p_{\ kl} \, \widetilde b_+^k \, \widetilde b_+^l \, \widetilde b_-^q \,\vert0\rangle_- \quad .
\ee
Here, $\eta_{klm}= \sqrt{-G} \epsilon_{klm}$ denotes the Levi-Civita tensor in $\beta$-space, and the first index  on $\eta^p_{\ kl}$
is moved by the Lorentzian metric $G^{pq}$ in $\beta$-space. The general solution \eqref{NF3sol} is parametrized by the (pseudo-)scalar $f$
and the (dualized) tensor $h_{pq}$. The latter tensor is {\it not} symmetric in its two indices and has, in general, nine
independent components. With the additional degree of freedom described by the scalar $f$, this means that the general $N_F=3$
solution a priori contains ten independent components (as befits its belonging to a 10-dimensional subspace of the $N_F=3$ level). 

It was found in  Ref. \cite{Damour:2014cba} that, far from all the walls, the general propagating solution at level $N_F=3$ simplifies because several
irreducible components among the ten generic ones either vanish or become related to each other. Specifically, in our
canonical chamber both the scalar $f$, and the antisymmetric part of the tensor $h_{pq}$, vanish far from the walls.
In addition, the remaining components, namely the six components of  the symmetric part $h_{(pq)}$ of $h_{pq}$,
can all be polynomially expressed in terms of the shifted momenta $\pi'_a$ according to a formula of the same type
as for the $N_F=2$ solution, i.e.
\begin{equation}
\label{asNF3}
h_{(pq)}^{\rm far-wall} = C_{N_F=3} e^{i \pi'_a \beta^a} e^{\varpi_a \beta^a} \left(\pi'_p \, \pi'_q + L_{pq}^k \, \pi'_k + m_{pq}\right) \, ,
\end{equation}
where $L_{pq}^k$ and  $m_{pq}$ are some fixed numerical coefficients (see Eqs. (19.32) and (19.33) in 
\cite{Damour:2014cba}, which are reproduced in Appendix A for the reader's convenience). [The coefficients $L_{pq}^k$ and  $m_{pq}$
describing the $N_F=3$ far-wall solutions are different from their $N_F=2$ analogs.]
Here the shifted (far-wall) momenta $\pi'_a$ satisfy the (non-tachyonic) mass-shell condition
\be \label{NF3massshell}
 G^{ab} \pi'_a  \pi'_b= -\mu^2_{N_F=3}= -\frac12 \quad .
\ee
The real exponential factor $e^{\varpi_a \beta^a} = e^{\beta^1+ \frac34 \beta^2+ \frac12 \beta^3}$ is the far-wall asymptotic form
of the rescaling factor $F(\beta)$, Eq. \eqref{F}.

We wish to generalize to the $N_F=3$ level the reflection laws \eqref{reflection1}, \eqref{reflection2}, \eqref{reflection3},
discussed above for the $N_F=2$ level. As before, these reflection laws will be obtained by matching the general
far-wall solution \eqref{asNF3} to three separate approximate susy solutions, obtained by considering, in turn, the various one-wall
cases where the solution propagates near each one of the three walls of our canonical chamber, i.e., the symmetric walls
$\alpha_{23}$ or $\alpha_{12}$, and the gravitational wall $\alpha_{11}$.
We will again find that the first reflection law \eqref{reflection1} is always satisfied, and we will compute the values of the
other scattering data, namely the $N_F=3$ global phase factor
\be\label{reflection2bis}
C_{N_F=3}^{\rm out} = e^{i \delta_{\alpha}^{\rm global}}  \, C_{N_F=3}^{\rm in}  \,,
\ee
and the reflection operator acting in the considered 10-dimensional subspace of the $N_F=3$ level, such that
\be\label{reflection3NF3}
|\Psi \rangle^{\rm out}_{{\bf 10}, N_F=3}= \mathcal{ R}_\alpha^{{\bf 10}, N_F=3}   |\Psi \rangle^{\rm in}_{{\bf 10}, N_F=3} \quad .
\ee
As in the $N_F=2$ case, we found that it is very useful to use (for each wall) the same wall-adapted basis as above to be able
both to solve the corresponding one-wall susy constraints, and to compute the scattering data. When working in some basis of
one forms, say $\alpha^{\widehat a}(\beta) = \alpha^{\widehat a}_p \beta^p$, we shall write the general solution \eqref{NF3sol}
in the form
\be \label{NF3solbasis}
|\Psi \rangle_{{\bf 10}, N_F=3} =f(\beta) \, |\eta \rangle +  h_{\widehat a \widehat b}(\beta) \,   | \alpha^{\widehat a} \alpha^{\widehat b} \rangle \, ,
\ee
where $ h_{\widehat a \widehat b} \equiv \alpha_{\widehat a}^p \alpha_{\widehat b}^q  h_{pq} $ are the components of the tensor 
$h_{pq}$ in the dual basis  ($\alpha_{\widehat a}^p  \alpha^{\widehat a}_q = \delta^p_q$ or, equivalently $\alpha_{\widehat a}^p \alpha^{\widehat b}_p= \delta_{\widehat a}^{\widehat b}$), and where we introduced the short-hand notation
\be \label{Bab}
| \alpha^{\widehat a} \alpha^{\widehat b} \rangle \equiv \alpha^{\widehat a}_p \, \alpha^{\widehat b}_q  \, | B^{pq} \rangle  \quad .
\ee
[One must also tensorially transform the Levi-Civita tensor,
and the metric.]

In the following subsections, we shall briefly summarize the main results of our analysis at the $N_F=3$ level.

\subsection{$N_F=3$ reflection on the symmetry wall $\alpha_{23}(\beta) = \beta^3- \beta^2$.}

We use the same basis of one forms as above, namely \eqref{basisNF2a23}, with \eqref{a23} and \eqref{uv}.
One decomposes the wavefunctions $f(\beta)$ and   $h_{\widehat a \widehat b}(\beta)$ entering the general $N_F=3$ solution \eqref{NF3solbasis} in the products
of the factor
\be
e^{(i\pi'_u-\frac 12) u+ (i \pi'_v +\frac 38) v} \,,
\ee
and of functions of $\alpha=\alpha_{23}(\beta)$. Here, the conserved shifted momenta $\pi'_u ,\pi'_v$ (which measure the momentum
parallel to the wall plane) must satisfy (when receding far from the wall) the $N_F=3$ mass-shell condition \eqref{NF3massshell}
which explicitly yields
\begin{equation}
2\,\pi^{\prime2}_\perp+4\,\pi'_u\pi'_v=  -\mu^2_{N_F=3}=-\frac 12 \, .
\end{equation}

 One then writes down the $N_F=3$ analogs of  equations \eqref{SAEqs2}, \eqref{delSig} and \eqref{compaS}  (written in terms of adapted-basis objects). The rank of the latter linear system is again found to be equal to 2. This means that the ten components of $f, h_{\widehat a \widehat b}$
can be expressed as linear combinations of only two of them. One also finds that the three antisymmetric components of 
$ h_{\widehat a \widehat b}$ must separately vanish. We could then express the seven remaining components,
 i.e. $f, h_{(\widehat a \widehat b)}$, in terms of two functions of $\alpha =\alpha_{23}(\beta) = \beta^3- \beta^2$,
 say $F(\alpha)$ and $G(\alpha)$. For instance,
 \begin{eqnarray} \label{hperpperp}
 h_{\perp \perp} &\propto& \sinh^{\frac38}(\alpha)\, G(\alpha) \, , \\ 
 h_{(\perp u)} &\propto& \sinh^{\frac38}(\alpha)\, F(\alpha)\, .
 \end{eqnarray}
 The $F$ and $G$ have  to satisfy the differential system
 \begin{eqnarray} \label{geq1020}
&&\partial_\alpha F +\frac 14\,\coth(\alpha)\, F+ G =0\label{feq1020}  \, ,\\
&&\partial_\alpha G +\frac 34 \,\coth(\alpha)\, G-\left(\frac 1{16}+\pi^{\prime 2}_\perp\right) F =0 \, ,
\end{eqnarray}
from which follows
\begin{equation}
\partial^2_\alpha  F +\coth(\alpha)\,\partial_\alpha F +\left(\frac 1{4}+\pi^{\prime 2}_\perp-\frac 1{16\,\sinh^2(\alpha)}\right) F =0\, .
\end{equation}
The general solution of this system is
\begin{eqnarray}\label{f1020}
F &=&c_+\,P^{+\frac 14}_{-\frac 12+i\,\pi'_\perp}[\cosh(\alpha)]+c_-\,P^{-\frac 14}_{-\frac 12+i\,\pi'_\perp}[\cosh(\alpha)]\, ,
\\
G &=&c_+\,\left(\frac 1{16}+\pi^{\prime 2}_\perp\right)\,P^{-\frac 34}_{-\frac 12+i\,\pi'_\perp}[\cosh(\alpha)]-c_-\,P^{+\frac 34}_{-\frac 12+i\,\pi'_\perp}[\cosh(\alpha)]\, .\label{g1020}
\end{eqnarray}
The solution for $f$ and $h_{(\widehat a \widehat b)}$ also involves the combination $ G + \frac12 \coth(\alpha) F$ which can
be shown to be equal to
\be
 G + \frac12 \coth(\alpha) F = c_+\,P^{+\frac 54}_{-\frac 12+i\,\pi'_\perp}[\cosh(\alpha)]-c_-\left(\frac9{16}+\pi^{\prime 2}_\perp\right)\,P^{-\frac 54}_{-\frac 12+i\,\pi'_\perp}[\cosh(\alpha)] \,.
\ee
As we see, there is a two-parameter family of solutions: (i) the $c_+$ family involving $P^{+\frac 54}_{-\frac 12+i\,\pi'_\perp}$,
$P^{+\frac 14}_{-\frac 12+i\,\pi'_\perp}$ and  $P^{-\frac 34}_{-\frac 12+i\,\pi'_\perp}$; and (ii) the
$c_-$ family involving $P^{-\frac 54}_{-\frac 12+i\,\pi'_\perp}$,
$P^{-\frac 14}_{-\frac 12+i\,\pi'_\perp}$ and  $P^{+\frac 34}_{-\frac 12+i\,\pi'_\perp}$.
Near the wall ($\alpha \to 0$), the Legendre functions behave like (we recall that $\alpha >0$) : 
\be
P^\mu_\nu(\cosh\alpha)\sim \frac 1{\Gamma(1-\mu)}\left(\frac{ \alpha }{2} \right)^{-\mu}. 
\ee
Though the above Legendre functions enter the solution after being
multiplied by $\sinh^{\frac38}(\alpha)$, the $c_+$ family of solutions will be singular at $\alpha=0$ in a non square integrable way.
We therefore exclude it, and retain only the $c_-$ family of solutions. [This family is mildly singular at $\alpha=0$ because
of the presence of $\sinh^{\frac38}(\alpha) \,P^{+\frac 34}_{-\frac 12+i\,\pi'_\perp}[\cosh(\alpha)]$. But the latter
mode is square integrable.]

Finally, defining (for $\mu= -\frac54, -\frac14, +\frac34$) the mode functions 
\be \label{hmu}
h^{\mu}(\beta) \equiv e^{(i\pi'_u-\frac 12) u+ (i \pi'_v +\frac 38) v} \sinh^{\frac38}(\alpha) \,P^{\mu}_{-\frac 12+i\,\pi'_\perp}[\cosh(\alpha)] \,,
\ee
we have been able to write the only regular solution of the susy constraints near the $\alpha_{23}$ symmetry wall
as a sum of the type
\be\label{mudecomp}
|\Psi \rangle_{{\bf 10}, N_F=3} = C_3 \sum_{\mu= -\frac54, -\frac14, +\frac34} N^i_\mu(\pi'_u,\pi'_v) h^{\mu}(\beta) |\mu , i\rangle \,.
\ee
Here, $i$ is a degeneracy index, which labels, for each value of $\mu$  various states associated with the same value of the order $\mu$
of the corresponding Legendre mode $h^{\mu}(\beta)$.

Parallely to the $N_F=2$ analysis above, there is again a direct link between the various mode states $|\mu , i\rangle$ 
and the spinorial operator $\widehat S_{23}^2$. Namely, the states $|\mu , i\rangle$ span, for each value of $\mu$
 (when the degeneracy index $i$ varies), the eigenspace of $\left[\widehat S_{23}^2\right]_{{\bf 10}, N_F=3}$
with eigenvalue $(2 \mu)^2$. More precisely, we have
\be \label{S23squared}
\left[\widehat S_{23}^2\right]_{{\bf 10}, N_F=3} |\mu , i\rangle =  (2 \mu)^2  |\mu , i\rangle \, {\rm for} \, i=1, \ldots g(\mu) \, ,
\ee
where the various degeneracies (which sum, as needed, to ten) are
\be
g\left(-\frac54\right)=1 \, ; \, g\left(-\frac14\right) = 5 \, ; \, g\left(+\frac34\right)= 4 \, .
\ee
Let us briefly indicate the structure of the various eigenstates, and how they are intimately linked to the basis adapted
to the considered wall $\alpha^{\perp} =\alpha_{23}$
\begin{eqnarray}
|-\frac54 \rangle &=& |u \, v \rangle + |v \, u \rangle -  | \perp \perp\rangle  -  |\eta \rangle   \, ,\\
|-\frac14 , i\rangle_{i=1,\ldots,5} &=&  | \perp u\rangle , | u \perp \rangle,   | \perp v\rangle , | v \perp \rangle, |\eta\rangle + |G\rangle \, ,\\
|+\frac34 ,  i\rangle_{i=1,\ldots,4} &=& | \perp \perp\rangle  -  |\eta\rangle \, , \, | u u\rangle, |v v\rangle, \, | u v\rangle- |v u\rangle  \, ,
\end{eqnarray}
where we used the notation \eqref{Bab}, together with the following shorthand for the trace state $|G\rangle \equiv G_{pq}| B^{pq}\rangle=  G_{\widehat a \widehat b} | \alpha^{\widehat a} \alpha^{\widehat b} \rangle = \frac12 \left(|\perp \perp \rangle + |uv\rangle  + |vu\rangle \right)$.

The possibility of expressing the solution of the susy constraints near a symmetry wall $\alpha_S$ as a combination of modes of the  type \eqref{hmu}
(involving Legendre functions $P^\mu_\nu(\cosh(\alpha))$) can be directly seen when considering the second-order 
equation (Hamiltonian constraint) which must be satisfied as a consequence of the first-order susy constraints. Indeed, the near-wall
form of the Hamiltonian constraint reads (with $ \hat \pi_a = - i \partial/\partial \beta^a$, and  $ |\Psi'(\beta)\rangle = F(\beta)^{-1}  |\Psi(\beta)\rangle$)
\be
 \left( G^{ab} \hat \pi_a  \hat \pi_b  + \mu^2_{N_F} + \frac12 \frac {\widehat S_{\alpha_S}^2-Id}{\,\sinh^2(\alpha_S)} \right) 
  |\Psi'(\beta)\rangle =0  \quad .
\ee
Decomposing the solution of this near-symmetry-wall second-order equation  in 
eigenspinors of the squared spin operator $\widehat S_{\alpha_S}^2$, one finds that the general solution pertaining to an eigenvalue $S^2$
of $\widehat S_{\alpha_S}^2$ is expressible in terms of the Legendre modes \eqref{hmu} for 
\be
\mu = \pm \frac{|S|}{2} \,.
\ee
As the eigenvalues $S^2$ (with multiplicities) of the squared spin operators at level $N_F=3$ are (for all symmetry walls)
$\left( \left( \frac52 \right)^2 \bigl\vert_1, \left( \frac12 \right)^2 \bigl\vert_5  , \left( \frac32 \right)^2 \bigl\vert_4 \right)$,
we recover the fact that the Legendre order $\mu$ can take the values 
$ \pm \frac54, \pm \frac14, \pm \frac34$. However, such an analysis based on the second-order equation alone cannot determine
which subset of indices $\mu$ belong to a given solution of the first-order susy constraints. Nor can they determine the subset
of indices belonging to a square-integrable solution, by contrast to a non square-integrable one. To determine that the one-parameter
family of square-integrable solutions of the susy constraints were associated with the set $\mu= \left\{ -\frac54, -\frac14, +\frac34 \right\}$
of indices we had to go through the more complicated analysis of the susy constraints sketched above.

Finally, we can extract from our analysis the scattering data for the $\alpha_{23}$ symmetry-wall reflection. The basic fact to be used is
the asymptotic decomposition of the Legendre function $ P^{\mu}_{\nu}$ given in Eq. \eqref{PmuAs} above. To determine the global phase
relating the incident far-wall amplitude $C_{N_F=3}$ to the reflected one, it is enough (as in the $N_F=2$ case) to consider the 
$\perp \perp$ component of the wave amplitude $h_{pq}$. [Indeed, we have checked that, for the $\alpha_{23}$ symmetry-wall reflection,
 the $N_F=3$ coefficient $ L_{\perp \perp}^{\perp}$
measuring the sensitivity of $h_{\perp \perp}^{\rm farwall}$, Eq. \eqref{asNF3}, to the sign of $\pi'_{\rm perp}$ vanishes]. We have exhibited in Eq. \eqref{hperpperp}, the fact that $ h_{\perp \perp}$ is proportional to $G(\alpha)$,
and therefore (for the square-integrable solution) to $P^{+\frac 34}_{-\frac 12+i\,\pi'_\perp}[\cosh(\alpha)]$. This shows that
the global phase factor is the one belonging to the value $\mu= + \frac34$. Using, the general result \eqref{deltamu}, we then get
\be
\left[ e^{i\,\delta_{\rm global}}\right]_{\alpha_{23}}=\frac{\Gamma(-\frac14 -i\,\pi'_\perp)\,\Gamma( i\,\pi'_\perp)}{\Gamma(-\frac14 +i\,\pi'_\perp)\,\Gamma(-i\,\pi'_\perp)} \quad .
\ee
In the WKB limit this yields
\be
\left[ e^{i\,\delta_{\rm global}}\right]_{\alpha_{23}}^{WKB} = e^{i \frac{\pi}{4}} \quad .
\ee
Moreover, the map between the incident spinor state and the reflected one is obtained by the reflection operator
\be \label{Ra23NF3}
\mathcal{ R}_{\alpha_{23}}^{{\bf 10}, N_F=3} = \frac{\Gamma[+i\pi'_\perp]\,\Gamma[\frac 12 -i\,\pi'_\perp- \frac12 \sqrt{\widehat S_{23}^2}_{{\bf 10}, N_F=3}]}{\Gamma[-i\pi'_\perp]\,\Gamma[\frac 12 +i\,\pi'_\perp- \frac12 \sqrt{\widehat S_{23}^2}_{{\bf 10}, N_F=3}]}\quad ,
\ee
which yields in the WKB limit
\be\label{Ra23NF3WKB}
\mathcal{ R}_{\alpha_{23}}^{{\bf 10}, N_F=3, WKB} = e^{- i \frac{\pi}{2}} e^{i \frac{\pi}{2} \sqrt{\widehat S_{23}^2}_{{\bf 10}, N_F=3}} \quad .
\ee
Here, $\sqrt{\widehat S_{23}^2}_{{\bf 10}, N_F=3}$ denotes an operator squareroot of ${\widehat S}_{23}^2{}_{{\bf 10}, N_F=3}$
which is {\it not} equal to its positive squareroot, but which is defined as
\be
\sqrt{\widehat S_{23}^2}_{{\bf 10}, N_F=3} \equiv 2 \, \widehat \mu \quad ,
\ee
by which we mean the following squareroot version of Eq. \eqref{S23squared}
\be
\sqrt{\widehat S_{23}^2}_{{\bf 10}, N_F=3} |\mu , i\rangle =  2 \, \mu  \, |\mu , i\rangle \, {\rm for} \, 2\mu=\left\{ -\frac52, -\frac12, +\frac32  \right\} \, {\rm and} \, i=1, \ldots g(\mu)  \,.
\ee

\subsection{$N_F=3$ reflection on the symmetry wall $\alpha_{12}(\beta) = \beta^2- \beta^1$.}

Concerning the reflection on the second symmetry wall of our canonical chamber, namely $\alpha_{12}(\beta) = \beta^2- \beta^1$,
the needed computations are very similar to the ones above, with, however, some significant differences. Though one would have expected
that a simple cyclic permutation would suffice to translate the results of the  $\alpha_{23}$ wall into results for the  $\alpha_{12}$
wall, there are some subtleties in intermediate results, linked to the fact that the explicit form of the susy constraints is not manifestly
cyclically symmetric. However, the end results are correctly obtained from a permutation $(231) \to (123)$.

We have already introduced above the basis adapted to the $\alpha_{12}(\beta) = \beta^2- \beta^1$ symmetry wall, namely Eqs.
\eqref{basisNF2a12}, \eqref{tildedbasis}. In terms of the frame components of the state \eqref{NF3solbasis}, there are
some simplifications because we found that the algebraic constraints on the state imply the vanishing not only (as before)
of the antisymmetric components of $h_{\widehat a \widehat b}$, but also the vanishing of the scalar $f$. This leaves us with
only six propagating components: $h_{(\widehat a \widehat b)}$. Again the adapted-frame decomposition of these
components is directly linked with eigenstates of the relevant squared spin operator, namely  $\left[\widehat S_{12}^2\right]_{{\bf 10}, N_F=3}$. The good  (square-integrable) modes are again of the form \eqref{hmu} with the corresponding $\mu$-decomposition
\eqref{mudecomp} of the solution. However, there is a difference in the link between each Legendre $P^\mu_\nu$ mode
and eigenspinors of $\left[\widehat S_{12}^2\right]_{{\bf 10}, N_F=3}$, with eigenvalues $ (2 \mu)^2$, as in Eq. \eqref{S23squared} above.
We have now, when considering a full basis of the 10-dimensional space, even if some coefficient modes vanish\footnote{The vanishing of such or such component depends on the choice of basis. What is important is that we were able to describe the exact solution of the susy constraints
within the ten-dimensional (half) $N_F=3$ state space.}
\begin{eqnarray}
|\mu=-\frac54 \rangle &=& | \perp \perp\rangle  ,\, \\
|\mu=-\frac14 , i\rangle_{i=1,\ldots,5} &=& |\eta \rangle \, ,\,  | u \perp \rangle \,, \,  |  \perp u \rangle \, , \, | v \perp \rangle \, ,\,  |  \perp v \rangle \, ,\\
|+\frac34 ,  i\rangle_{i=1,\ldots,4} &=&  | u u\rangle, |v v\rangle, \, | u v\rangle,  |v u\rangle \quad .
\end{eqnarray}

Let us only exhibit here, for illustration, the form of the $\perp \perp $ mode:
\be
h_{\perp \perp} \propto \sinh^{\frac38}(\alpha) P^{-\frac 54}_{-\frac 12+i\,\pi'_\perp}[\cosh(\alpha)] \quad .
\ee
Contrary to the $N_F=2$ case, $h_{\perp \perp}^{\rm far wall}$ is sensitive to the sign of $\pi'_\perp$ (i.e. the projected coefficient 
$L_{\perp \perp}^{\perp}$ does not vanish). However, the other projections $L_{U V}^{\perp}$ of $L^k_{pq}$ do vanish,
so that the farwall parallel-parallel components of $h_{pq}$ are insensitive to the sign of $\pi'_\perp$.
Finally, the global phase factor for the $N_F = 3$ reflection on the $\alpha_{12}$ symmetry wall is given by the behavior of
the $\mu=\frac34$ mode, i.e.
\be
\left[ e^{i\,\delta_{\rm global}}\right]_{\alpha_{12}}=\frac{\Gamma(-\frac14 -i\,\pi'_\perp)\,\Gamma( i\,\pi'_\perp)}{\Gamma(-\frac14 +i\,\pi'_\perp)\,\Gamma(-i\,\pi'_\perp)}  \, ,
\ee
with WKB limit:
\be
\left[ e^{i\,\delta_{\rm global}}\right]_{\alpha_{12}}^{WKB} = e^{i \frac{\pi}{4}} \quad .
\ee
The corresponding reflection operator reads
\be \label{Ra12NF3}
\mathcal{ R}_{\alpha_{12}}^{{\bf 10}, N_F=3} = \frac{\Gamma[+i\pi'_\perp]\,\Gamma[\frac 12 -i\,\pi'_\perp- \frac12 \sqrt{\widehat S_{12}^2}_{{\bf 10}, N_F=3}]}{\Gamma[-i\pi'_\perp]\,\Gamma[\frac 12 +i\,\pi'_\perp- \frac12 \sqrt{\widehat S_{12}^2}_{{\bf 10}, N_F=3}]}\qquad ,
\ee
(where ${\sqrt{\widehat S_{12}^2}}_{{\bf 10}, N_F=3}$ is again defined as being $ 2 \widehat \mu$) which yields in the WKB limit
\be\label{Ra12NF3WKB}
\mathcal{ R}_{\alpha_{12}}^{{\bf 10}, N_F=3, WKB} = e^{- i \frac{\pi}{2}} e^{i \frac{\pi}{2} \sqrt{\widehat S_{12}^2}_{{\bf 10}, N_F=3}} \,.
\ee

\subsection{Reflection at level $N_F = 3$ on the  gravitational wall $\alpha_{11}(\beta) = 2 \beta^1$. } \label{GW1020}

When considering the reflection  on the  gravitational wall $\alpha_{11}(\beta) = 2 \beta^1$ of a  $N_F = 3$ solution we use
the same adapted basis as in our corresponding $N_F=2$ analysis, namely \eqref{gbasis}. 
Instead of the Legendre-like mode functions \eqref{hmu}, we will have Whittaker-like mode functions, 
${\mathcal U}_{\mu^2} (\beta ; j)$, as defined in Eq. \eqref{Uj}. The only difference is that the squared mass value $\mu^2$ 
labelling these modes must now be taken to be $\mu^2_{N_F=3} = + \frac12$ (instead of $\mu^2_{N_F=2} = - \frac38$).
Actually, the value of $\mu^2$ only enters indirectly in the expression of ${\mathcal U}_{\mu^2} (\beta ; j)$ via a modified
link between the shifted parallel momenta $\pi'_u ,\pi'_v$ and $\pi'_\perp$. In the present case, this explicit link reads: 
$\pi'_\perp=\sqrt{\pi'_u\pi'_v- \frac14}$.

In the case of symmetry walls, we were decomposing the state $|\Psi \rangle_{{\bf 10}, N_F=3}$ into eigenstates of the
squared spin operator ${\widehat S}_{23}^2{}_{{\bf 10}, N_F=3}$(labelled by $\mu$ with $ (2 \mu)^2= S^2$ measuring the eigenvalues of ${\widehat S}_{23}^2{}_{{\bf 10}, N_F=3}$), as in Eq. \eqref{mudecomp}. Here, we shall decompose $|\Psi \rangle_{{\bf 10}, N_F=3}$ into eigenstates of the operator $\widehat J_{11}$ (with eigenvalues denoted $j$), according to
 \be\label{jdecomp}
|\Psi \rangle_{{\bf 10}, N_F=3} = C_3 \sum_{j=-2,0,2} N^i_j(\pi'_u,\pi'_v)  {\mathcal U}_{\frac12} (\beta ; j) |j , i\rangle  \,.
\ee
At level 3, the eigenvalues $j$ (with their degeneracies labelled above by $i$) of $\left[\widehat J_{11}\right]_{{\bf 10}, N_F=3}$
are $\left(+2\vert_2,\, 0\vert_6,\,-2\vert_2\right)$. More importantly, the eigenspinors corresponding to these eigenvalues
are directly linked with objects naturally constructed within our present adapted basis. Namely, we have (using the notation \eqref{Bab},
now applied to our new adapted basis)
\begin{eqnarray}
|j=2 ,i \rangle_{i=1,2} &=& |u \, \perp \rangle , |v \, \perp \rangle  \, ,\\
|j=0 , i\rangle_{i=1,\ldots,6} &=&  | \perp \perp\rangle , | u u\rangle,   | v v\rangle , | u v \rangle, |v u\rangle,|\eta\rangle  \, ,\\
|j=-2,  i\rangle_{i=1,2} &=& |\perp , \,  u\rangle ,  |\perp , \,  v\rangle \, .
\end{eqnarray}

Let us only cite the form of our final result, namely the expression of all the components $h_{\widehat a \widehat b}(\beta)$
(modulo an overall factor that we omit)
of the main $N_F=3$ polarization tensor $h_{pq}$ along our adapted basis. [The scalar polarization $f$ happens to vanish,
as well as  $h_{uv} - h_{vu}$.]
\begin{eqnarray} 
&&h_{u\perp}= {-i\,\pi'_u}\ {\cal U}_{\frac 12}(\beta,+2)  \, , \nonumber \\
&&h_{v\perp} =-\frac {i\,\pi'_u\,\pi'_v}{(\pi'_u-i/2)}\ {\cal U}_{\frac 12}(\beta,+2)  \, , \nonumber \\
&&h_{\perp u}= \frac{i}{\pi'_v} \ {\cal U}_{\frac 12}(\beta,-2) \, , \nonumber  \\
&&h_{\perp v} =\frac {i}{(\pi'_u-i/2)}\ {\cal U}_{\frac 12}(\beta,-2) \, , \nonumber \\
&&h_{\perp\perp} = \frac12\frac{(2\,\pi'_u\,\pi'_v- i\,\pi'_u-1/2)}{\pi'_v( \pi'_u-i/2)} {\cal U}_{\frac 12}(\beta,0) \, , \nonumber  \\
&&h_{uu} =\frac {(\pi'_u-i/2)}{2\,\pi'_v}\ {\cal U}_{\frac 12}(\beta,0) \, , \nonumber \\
&&h_{vv} =\frac { \pi^{\prime 2}_v+ i\,(\pi'_u-\pi'_v)+1/2}{2\,\pi'_v\,(\pi'_u-i/2)}\ {\cal U}_{\frac 12}(\beta,0) \, , \nonumber \\
&& \frac12 \left( h_{uv} + h_{vu} \right)=\frac12 \ {\cal U}_{\frac 12}(\beta,0) \, .
\label{sola11NF3}
\end{eqnarray}

Using the asymptotic behaviour of the Whittaker modes [see Eq. \eqref{asympU}], we deduce the reflection laws on the gravitational wall $2\,\beta^1=0$. We checked that (because, for the present case, $L_{\perp\perp}^{\perp}=0$) the global phase is read off the $ h_{\perp\perp}$ expression (involving $j=0$) and reads
\begin{equation} \label{deltajNF3}
e^{i\delta_{\alpha_{11}}^{\rm global}} = \frac{\Gamma \left[ \frac{1}2 - i \,\pi'_\perp \right] \Gamma \left[ i\,2\,\pi'_\perp \right]}{\Gamma \left[ \frac{1}2 + i \,\pi'_\perp \right] \Gamma \left[ -i\,2\,\pi'_\perp \right]} \quad .
\end{equation}
As before it is energy-dependent, and has no limit as $\pi'_\perp  \to +\infty$.

The  reflection operator  against the $\alpha_{11}$ gravitational wall (acting in Hilbert space and transforming the incident
state into the reflected one) is given by the following operatorial expression
\be\label{Ra11NF3}
\mathcal{ R}_{\alpha_{11}}^{{\bf 10}, N_F=3} = \frac{\Gamma \left[ \frac{1+\widehat J_{11}}2 - i \,\pi'_\perp \right] \Gamma \left[ i\,2\,\pi'_\perp \right]}{\Gamma \left[ \frac{1+\widehat J_{11}}2 + i \,\pi'_\perp \right] \Gamma \left[ -i\,2\,\pi'_\perp \right]} \, .
\ee
In the large momentum (WKB) limit, this yields
\be
\mathcal{ R}_{\alpha_{11}}^{{\bf 10}, N_F=3} =e^{i \delta_0(\pi'_\perp)} e^{- i \frac{\pi}{2}}  e^{- i \frac{\pi}{2} \widehat J_{11}} \, ,
\ee
where $\delta_0(\pi'_\perp) = 2 \pi'_\perp \ln( 4 \pi'_\perp/e)$.
These are formally the same expressions as at level $N_F=2$, but, here, $\widehat J_{11}$ denotes the endomorphism of
the 10-dimensional $N_F=3$ subspace in which we are working.

Let us note that the solution \eqref{sola11NF3} contains more excited components than the previous symmetry-wall $N_F=3$
solutions. In particular, the antisymmetric components  $h_{u\perp}- h_{\perp u}$ and $h_{v\perp}- h_{\perp v}$ do not vanish,
while they vanished before. However, using the asymptotic behavior, Eq. \eqref{asympU}, of the 
relevant functions $ {\cal U}_{\frac 12}(\beta ,\pm 2)$, one finds that their leading-order asymptotic approximations (as $\beta^1 \to +\infty$)
are exactly  proportional to each other:
\begin{equation}
 {\cal U}^{\rm asympt}_{\frac 12}(\beta,+2)  = -\frac 1{\pi'_u\,\pi'_v}\,  {\cal U}^{\rm asympt}_{\frac 12}(\beta,-2)\, .
 \end{equation}
 Inserting this asymptotic relation in Eqs. \eqref{sola11NF3} one finds that the antisymmetric components $h_{u\perp}- h_{\perp u}$ and $h_{v\perp}- h_{\perp v}$ vanish far from the gravitational wall (in keeping with the far-wall analysis of Ref. \cite{Damour:2014cba}).
 
 \section{Hidden Kac-Moody structure of the spinor reflection operators}
 
 Let us consider the WKB limit of the reflection operators $\mathcal{ R}_{\alpha}^{ \rm rep}$ that map the incident
 spinor states $|\Psi \rangle^{\rm in}$ to the reflected ones $|\Psi \rangle^{\rm out}$. These spinor reflection operators
 depend both on the considered reflection wall form $\alpha(\beta)$ and on the representation space, say $V_{\rm rep}$, in which
 lives the  considered incident and reflected quantum states. More precisely, we derived above two different triplets of
 such reflection operators: (i) one triplet associated with the reflection (on the three potential walls of our canonical billiard chamber) of the propagating quantum susy states  at level $N_F=2$, which live in a 6-dimensional representation; and (ii) a second triplet associated
 with the reflection (on the same three bounding walls)  of the propagating quantum susy states  at level $N_F=3$, which live in a 10-dimensional representation.  In the WKB limit (and after factorization of the classical, energy-dependent part of the gravitational-wall reflection,
  $\delta_0(\pi'_\perp) = 2 \pi'_\perp \ln( 4 \pi'_\perp/e)$),
 we found the following operatorial expressions for these two triplets of reflection operators:
 \begin{eqnarray} \label{RNF2}
\mathcal{ R}_{\alpha_{23}}^{{\bf 6}, N_F=2, WKB} &=& e^{-\frac{ i\pi}{2}} \, e^{ \pm \frac{ i\pi}{2} |\widehat S_{23}|_{{\bf 6}, N_F=2} } \,,\nonumber \\
\mathcal{ R}_{\alpha_{12}}^{{\bf 6}, N_F=2, WKB} &=& e^{-\frac{ i\pi}{2}}
e^{ \pm  \frac{ i\pi}{2} |\widehat S_{12}|_{{\bf 6}, N_F=2}}  \,,\nonumber \\
\mathcal{ R}_{\alpha_{11}}^{{\bf 6}, N_F=2} &=& e^{- i \frac{\pi}{2}}  e^{- i \frac{\pi}{2} {\widehat J}_{11}{}^{{\bf 6}, N_F=2}} \,,
\end{eqnarray}
where we recall that $ |\widehat S_{23}|_{{\bf 6}, N_F=2} $ and  $  |\widehat S_{12}|_{{\bf 6}, N_F=2} $ were defined as
the {\it positive} squareroots of the corresponding squared spin operators $\widehat S_{23}^2$, $\widehat S_{12}^2$, which are
both endomorphisms of the 6-dimensional subspace ${\mathbb H}_{(1,1)_S}$ of the $N_F=2$ level. The ``gravitational" operator 
$\widehat J_{11}$ is also an endomorphism of ${\mathbb H}_{(1,1)_S}$. [See, e.g., the second Table in Appendix B of \cite{Damour:2014cba}.]

 The corresponding results for the reflection operators in the 10-dimensional subspace of the $N_F=3$ level where live the propagating
 quantum states read:
  \begin{eqnarray}\label{RNF3}
\mathcal{ R}_{\alpha_{23}}^{{\bf 10}, N_F=3, WKB} &=& e^{- i \frac{\pi}{2}} e^{i \frac{\pi}{2} \sqrt{\widehat S_{23}^2}_{{\bf 10}, N_F=3}} \,, \nonumber \\
\mathcal{ R}_{\alpha_{12}}^{{\bf 10}, N_F=3, WKB} &=& e^{- i \frac{\pi}{2}} e^{i \frac{\pi}{2} \sqrt{\widehat S_{12}^2}_{{\bf 10}, N_F=3}} \,, \nonumber \\
\mathcal{ R}_{\alpha_{11}}^{{\bf 10}, N_F=3} &=& e^{- i \frac{\pi}{2}}  e^{- i \frac{\pi}{2} \widehat J_{11}{}^{{\bf 10}, N_F=3}}\,.
\end{eqnarray}

Here, there is a crucial difference in the way the squareroots of the squared-spin operators are defined. We recall that both squared-spin
operators have eigenvalues $ (2\mu)^2=\left\{ (\frac52)^2, (\frac12)^2, (\frac32)^2  \right\}$. The squareroot 
operators $\sqrt{\widehat S_{ab}^2}{}_{{\bf 10}, N_F=3}$ are defined as having the eigenvalues
$ 2\mu=\left\{ -\frac52, -\frac12, +\frac32  \right\}$ on the corresponding eigenspaces of ${\widehat S}_{ab}^2{}_{{\bf 10}, N_F=3}$.
This sign pattern is such that the corresponding, successive values of the Legendre order $\mu$, namely 
$\left\{ -\frac54, -\frac14, +\frac34  \right\}$  differ by 1 (so as to correspond to the regular solution of the first-order susy
constraints).

Let us emphasize that the results above for the reflection operators have resulted from a purely {\it dynamical} computation
within supergravity. However, a remarkable fact is that the end results of these supergravity calculations can be expressed in
terms of mathematical objects having a (hyperbolic) Kac-Moody meaning. More precisely, we are going to show that the
two triplets of spinorial reflection operators satisfy some relations that are related to a spin-extension of the Weyl group of the
rank-3 hyperbolic Kac-Moody algebra $AE_3$. The notion of spin-extended Weyl group was introduced, 
within the use of specific representations of the maximally compact subalgebra $K[AE_3]$ of
$AE_3$ (and $K[E_{10}] \subset E_{10}$), in Ref. \cite{Damour:2009zc}. More precisely, Ref. \cite{Damour:2009zc} studied
the one-wall reflection laws of the classical, Grassmann-valued gravitino field $\psi$, in the case where, near each potential wall
(with bosonic potential $\propto e^{- 2\alpha(\beta)}$),
the coupling of the gravitino is  also Toda-like and $\propto e^{- \alpha(\beta)}$, so that the law of evolution
of $\psi$ near each separate wall reads
\be
\partial_t \psi \approx i \, e^{- \alpha(\beta)} \Pi_{\alpha}  J_{\alpha} \psi  \,,
\ee
where $\Pi_{\alpha} $ is a conserved momentum.

Under these assumptions, Ref. \cite{Damour:2009zc} found that the transformation linking the incident value
of the Grassmann-valued $\psi$ to its reflected value was given by a classical, fermionic reflection operator of the form
\be\label{RalphaG}
\mathcal{ R}_{\alpha}^{\rm classical} = e^{ i  \frac{\pi}{2} \varepsilon_{\alpha}  J_{\alpha}} \,,
\ee
where $\varepsilon_{\alpha} = \pm$ denotes the sign of the momentum $\Pi_{\alpha}$.
Here, $ J_{\alpha}$ is a matrix acting on the representation space defined by a classical homogeneous gravitino.
In the case of  Ref. \cite{Damour:2009zc}, this was (when considering 4-dimensional supergravity) a 12-dimensional space in which live the twelve components of a Majorana 
(spatial) gravitino $\psi^i_A$, with $i=1,2,3$ (spatial index) and $A=1,2,3,4$ (Majorana spinor index). This 12-dimensional
representation is (essentially) equivalent to the direct sum of the two (complex-conjugated)
$6$-dimensional complex representations that live at levels $N_F=1$ and $N_F=5$ within our 64-dimensional quantized-gravitino
Hilbert space. [In view of the hidden, but crucial, importance of the existence of such finite-dimensional representations, 
we briefly discuss in Appendix B the structure of some of the low-dimensional representations of $K[AE_3]$.]

Motivated by these physical findings, a mathematical definition of spin-extended Weyl groups (for general Kac-Moody algebras)
was then implemented (as part of the definition of spin-covers of  maximal compact Kac-Moody subgroups of the $K[AE_3]$ type)
and studied in Ref. \cite{Koehl}.

 Ref. \cite{Damour:2009zc}  showed that the reflection operators, say $r_i^G= \mathcal{ R}_{\alpha_i}^{\rm classical}$, describing the Grassmanian scattering on the dominant potential walls (labelled by the index $i=1, \cdots, {\rm rank}$) of the cosmological 
 supergravity billiards (both in dimension $D=11$ and in $D=4$) satisfied some spinorial generalization of the usual Coxeter relations 
 \footnote{In the notation of Eqs. \eqref{r8}, \eqref{braid} below, the usual Coxeter relations defining the Weyl group, i.e. the group
 generated by geometrical reflections in the simple-root hyperplanes in Cartan space are: $ r_i^2=1$ and the braid relations \eqref{braid}.}
 satisfied by the corresponding Weyl-group generators. [We recall that a basic finding of cosmological billiards \cite{Damour:2002et} is that the gravity-defined
 billiard chamber coincides with the Weyl chamber of some corresponding Kac-Moody algebra.] The (Grassman-supergravity-based)
 spin-extended Weyl group was then defined as the infinite, discrete matrix group generated by the $r_i^G$'s. [Here, the index $i$ labels
 the nodes of the Dynkin diagram, corresponding to the simple roots of a Kac-Moody algebra, and to the dominant walls of the supergravity dynamics.]
 The generalized Coxeter relations satisfied by the
 Grassmanian reflection operators $r_i^G$ can be written as\footnote{Here, following standard mathematical lore \cite{KacPeterson85}, we rewrite the relations
 written in  Ref. \cite{Damour:2009zc} in a form that only involves the multiplicative identity, rather than the ``minus identity operator" 
 used there when dealing with concrete, matrix forms of the $r_i^G$'s.}
 \be \label{r8}
 r_i^8=1;
 \ee 
 \be\label{braid}
 r_i r_j r_i \cdots = r_j r_i r_j \cdots \, {\rm with} \, m_{ij} \, {\rm factors \, on \, each\, side} \, .
 \ee
 Here, $i$, and $j$, with $i\neq j$ (which includes both $i<j$ and $i>j$), are labels for the nodes of the  Dynkin diagram of the considered Kac-Moody group.
 The positive integers $m_{ij}$ entering the ``braid relations" \eqref{braid} are defined from the corresponding values of the nondiagonal elements of the Cartan matrix $a_{ij}$ (which are supposed to be negative integers, while $a_{ii}=2$).
Namely (see \cite{KacPeterson85})
\be
m_{ij} = \left\{ 2,3,4,6,0 \right\} \, \, {\rm if} \, \, a_{ij}a_{ji}=  \left\{ 0,1,2,3, \geq 4 \right\} \, \,( {\rm respectively }) \, .
\ee
In addition to the generalized Coxeter relations, \eqref{r8}, \eqref{braid}, Ref. \cite{Damour:2009zc} had found that the squared
Grassman reflection operators ${(r_i^G)}{}^2$ had simple properties. Namely, they generated a finite-dimensional, normal subgroup of the 
corresponding (Grassman-based) spin-extended Weyl group.

According to the mathematical definition of Ref.\cite{Koehl}, the spin-extended Weyl group of a Kac-Moody algebra with Dynkin
diagram $\Pi$ is a discrete subgroup of a spin cover of the maximally compact Kac-Moody subgroup $K[\Pi]$
that is generated by elements of order eight (involving the polar angle $\frac{\pi}{4}$). This mathematically-so-defined spin-extended
Weyl group can also be charaterized by generators and relations. Namely, its (abstract) generators $r_i$ satisfy not only
the generalized Coxeter relations above \eqref{r8}, \eqref{braid}, but also the following ones:
\be \label{r2}
\, r_j^{-1} r_i^2 r_j = r_i^2 r_j^{2n_{ij}} \, ,
\ee
where, as above $i\neq j$, and where the positive integers $n_{ij}$ are defined from the corresponding values of the 
nondiagonal elements of the Cartan matrix $a_{ij}$ via
\be \label{defn}
n_{ij} = 0 \, ({\rm respectively} \, =1) \, {\rm if}\,  a_{ij} \, {\rm is \, even} \,( {\rm resp. \, odd}) \, .
\ee
The additional (non-Coxeter-like) relations \eqref{r2} are the same as those that enter the Tits-Kac-Peterson \cite{KacPeterson85}
extension of the Weyl group (generated by elements of order four: $t_i^4=1$). Their origin is not clear to us, and we shall see below
that the quantum-motivated reflection operators that have appeared in our dynamical study above, namely \eqref{RNF2} and \eqref{RNF2},
satisfy the generalized Coxeter relations \eqref{r8}, \eqref{braid}, but satisfy a phase-modified form of the (non-Coxeter-like) relations \eqref{r2}.

The Kac-Moody algebra that (in view of previous works) we expect to be relevant to our
present dynamical study is $AE_3$, and its Dynkin diagram is
\be \label{dynkin}
\xymatrix{
{ \underset{J_{11}}{ \bullet}\!\!\!} \ar@{<=>}[r] &{ \!\!\!\underset{S_{12}}{ \bullet} \!\!\!}   \ar@{-}[r] &{{\!\!\! \underset{S_{23}}{ \bullet} } }
}
\ee
Here, we use the labelling: $ (1,2,3) =  (J_{11}, S_{12}, S_{23})$. The two arrows and the double line
between nodes 1 and 2 mean that $a_{12} = a_{21}=-2$, while the single line between nodes 2 and 3 mean that $a_{23} = a_{32}=-1$.
Finally, $a_{13} = a_{31}=0$. As a consequence, the relevant values of the integers $n_{ij}$ and $m_{ij}$ to be used in 
Eqs. \eqref{r8}, \eqref{braid}, and \eqref{r2}, are:
\be
n_{12}=n_{21}=0\, ; \, m_{12}=m_{21}=0 \,; \, n_{23}=n_{32}=1\,; \,m_{23}=m_{32}=3 \,; \, n_{13}=n_{31}=0 \,; \,m_{13}=m_{31}=2  \, .
\ee
The three relations $r_i^8=1$, Eq. \eqref{r8}, are satisfied for each one of our triplets of reflection operators \eqref{RNF2}, \eqref{RNF3}.
[This is clear without calculation because the eigenvalues of all our reflection operators are $e^{i k \frac{\pi}{4}}$ for some integer $k$.]
By explicit (matrix) calculations, we have verified that the $AE_3$ braid relations \eqref{braid}, namely
\be
r_2 r_3 r_2=  r_3 r_2 r_3 \, ; \, r_1 r_3= r_3 r_1
\ee
(note that $m_{12}=m_{21}=0$ so that there are no braid relations between the nodes $J_{11}$ and $S_{12}$) are also
satisfied by our two triplets of reflection operators \eqref{RNF2}, \eqref{RNF3}.

 Concerning the non-Coxeterlike relations \eqref{r2}, let us first emphasize that we view them as expressing constraints
 on the sub-group generated by the squared operators $r_i^2$. As in Ref. \cite{Damour:2009zc}, we looked directly
 at the values taken (within the two matrix representations that we are considering) by the squares of our two triplets
 of generators \eqref{RNF2}, \eqref{RNF3}. We found that they have extremely simple values; namely they only differ
 from the identity matrix by some simple phase factors, namely
  \begin{eqnarray} \label{sqRNF2}
\left( \mathcal{ R}_{\alpha_{23}}^{{\bf 6}, N_F=2, WKB} \right)^2 &=&  \, - Id_{\bf 6}= e^{i \pi} Id_{\bf 6}  \,,\nonumber \\
\left(\mathcal{ R}_{\alpha_{12}}^{{\bf 6}, N_F=2, WKB} \right)^2 &=& \, - Id_{\bf 6} = e^{i \pi} Id_{\bf 6} \,, \nonumber \\
\left(\mathcal{ R}_{\alpha_{11}}^{{\bf 6}, N_F=2, WKB} \right)^2 &=& \, e^{- i\frac{\pi}{2} } Id_{\bf 6}  \,,
\end{eqnarray}
and
 \begin{eqnarray} \label{sqRNF3}
\left( \mathcal{ R}_{\alpha_{23}}^{{\bf 10}, N_F=3, WKB} \right)^2 &=&  \, e^{i \frac{ \pi}{2} } Id_{\bf 10} \,, \nonumber \\
\left(\mathcal{ R}_{\alpha_{12}}^{{\bf 10}, N_F=3, WKB} \right)^2 &=& \, e^{i \frac{ \pi}{2} } Id_{\bf 10} \,,\nonumber \\
\left(\mathcal{ R}_{\alpha_{11}}^{{\bf 10}, N_F=3, WKB} \right)^2 &=& \, - Id_{\bf 10}= e^{i \pi} Id_{\bf 10} \,.
\end{eqnarray}

 In both cases the subgroup generated by the squared reflection operators is central (i.e. commutes with everything else)
 and isomorphic to the multiplicative group of order four generated by $e^{i  \frac{\pi}{2}}$.
 
 Finally, in view of the simple results \eqref{sqRNF2}, \eqref{sqRNF3}, it is a simple matter to see whether the non-Coxeterlike relations \eqref{r2}
 are satisfied or not. One can easily see that, with the precise definitions \eqref{RNF2}, \eqref{RNF2}, they are not satisfied as
 written. However, they are satisfied modulo the inclusion of additional phase factors in the relations \eqref{r2}. The latter
 phase factors can be easily reabsorbed in suitable redefinitions of the basic reflection operators. For instance, if we
 had defined, at level $N_F=2$ (with an arbitrary integer $n$ in the third line)
 \begin{eqnarray} \label{RNF2new}
\mathcal{ R}_{\alpha_{23}}^{{\bf 6}, N_F=2, WKB, new} &=&  \, e^{ \pm \frac{ i\pi}{2} |\widehat S_{23}|_{{\bf 6}, N_F=2} } \,,\nonumber \\
\mathcal{ R}_{\alpha_{12}}^{{\bf 6}, N_F=2, WKB, new} &=& 
e^{ \pm  \frac{ i\pi}{2} |\widehat S_{12}|_{{\bf 6}, N_F=2}}  \,,\nonumber \\
\mathcal{ R}_{\alpha_{11}}^{{\bf 6}, N_F=2,new} &=& e^{ i n \frac{\pi}{4}}  e^{- i \frac{\pi}{2} {\widehat J}_{11}{}^{{\bf 6}, N_F=2}} \,,
\end{eqnarray}
and, at level $N_F=3$,
  \begin{eqnarray}\label{RNF3new}
\mathcal{ R}_{\alpha_{23}}^{{\bf 10}, N_F=3, WKB} &=& e^{- 3 i \frac{\pi}{4}} e^{i \frac{\pi}{2} \sqrt{\widehat S_{23}^2}_{{\bf 10}, N_F=3}} \,,\nonumber \\
\mathcal{ R}_{\alpha_{12}}^{{\bf 10}, N_F=3, WKB} &=& e^{- 3 i \frac{\pi}{4}} e^{i \frac{\pi}{2} \sqrt{\widehat S_{12}^2}_{{\bf 10}, N_F=3}} \,, \nonumber \\
\mathcal{ R}_{\alpha_{11}}^{{\bf 10}, N_F=3} &=& e^{ i n \frac{\pi}{4}}  e^{- i \frac{\pi}{2} \widehat J_{11}{}^{{\bf 10}, N_F=3}} \,,
\end{eqnarray}
these two new triplets of operators would satisfy all the relations \eqref{r8}, \eqref{braid}, {\it and} \eqref{r2}.
In that case, the corresponding squared operators are simply equal to unity (for an appropriate choice of $n$). 

Let us also mention in passing that if we define, within the full sixty-four-dimensional
spinorial space which gathers all the fermionic levels (from $N_F=0$ to $N_F=6$) the quantum analogs of the
Grassmann-motivated operators defined in Ref. \cite{Damour:2009zc}, namely
 \begin{eqnarray}\label{R64}
\mathcal{ R}_{\alpha_{23}}^{{\bf 64}} &=&  e^{- i \frac{\pi}{2} {\widehat S}_{23}^{\, {\bf 64}}} \,,\nonumber \\
\mathcal{ R}_{\alpha_{12}}^{{\bf 64}} &=&   e^{- i \frac{\pi}{2} {\widehat S}_{12}^{\, {\bf 64}}} \,, \nonumber \\
\mathcal{ R}_{\alpha_{11}}^{{\bf 64}} &=& e^{- i \frac{\pi}{2} {\widehat J}_{11}^{\, {\bf 64}}} \,,
 \end{eqnarray}
the latter reflection operators satisfy all the relations \eqref{r8}, \eqref{braid},  and \eqref{r2}.

\section{Conclusions}

We solved the susy constraints \eqref{susy} of the supersymmetric Bianchi IX model in the one-wall approximation, i.e.
taking into account one potential wall at a time. This allowed us to derive the quantum laws of reflection of the wave function
of the universe  $|\Psi(\beta) \rangle$ during its chaotic evolution near a big crunch singularity, i.e. in the domain of large
(positive) values of the three squashing parameters $\beta^1, \beta^2, \beta^3$ (considered in the symmetry chamber $\beta^1<\beta^2 <\beta^3$).
Our analysis could limit itself to two subspaces of the total 64-dimensional fermionic state space because we had shown
in previous work that propagating states only exist in subspaces of the fermion levels $N_F=2$, $N_F=3$ and $N_F=4$.
In addition, given the symmetry between the $N_F=2$ and the $N_F=4$ levels, and the self-symmetry of the 
$N_F=3$ level, and in view of the special structure of the propagating states, it was enough to work (separately) 
in a 6-dimensional subspace of the $N_F=2$ level, and in a 10-dimensional half of the $N_F=3$ level.

Our main results are contained in Eqs. \eqref{Ra23NF2},  \eqref{Ra11NF2}, \eqref{Ra23NF3}, \eqref{Ra12NF3}, \eqref{Ra11NF3}, and are summarized (in the small-wavelength limit, which allows one to highlight
their structure) in the reflection operators \eqref{RNF2}, \eqref{RNF3}. We remarkably found that 
the latter, purely dynamically-defined, reflection operators
satisfy generalized Coxeter relations which define a type of spinorial extension of the Weyl group of the rank-3 hyperbolic Kac-Moody
algebra $AE_3$. More precisely, we found that our dynamical reflection operators satisfy the generalized Coxeter relations
\eqref{r8} and \eqref{braid} associated with the Dynkin diagram \eqref{dynkin} of $AE_3$, and selected in Ref. \cite{Damour:2009zc}
(in a slightly different form) as characteristic of a spin-extension of the Weyl group. We also found that
the squares of our dynamical reflection operators commute with all the reflection operators. In addition, some phase-modified
versions of the reflection operators, see Eqs. \eqref{RNF2new}, \eqref{RNF3new} satisfy the relations \eqref{r2}
that are part of the defining relations of the mathematically-defined spin-extended Weyl group of Ref. \cite{Koehl}.
The fact that our dynamically-defined spinorial reflection operators satisfy relations that appear as being partly 
more general than those of Ref. \cite{Koehl} (though only modulo some extra phase factors, Eqs. \eqref{RNF2new}, \eqref{RNF3new})
might suggest the need to define more general spin-covers than those mathematically constructed in Ref. \cite{Koehl}.
Anyway, independently of such an eventual generalization, let us repeat that our findings provide a new evidence for
the existence of hidden Kac-Moody structures in supergravity. In particular, our results have gone beyond previous
related evidence for Kac-Moody structures in two directions: (i) we quantized the gravitino degrees of freedom instead of
treating $\psi_{\mu}$ as a classical, Grassmann-valued object, and (ii) in our quantum treatment the symmetry walls
necessarily involved operators quartic in fermions (through the squared spin operators $\widehat S_{12}^2, \widehat S_{23}^2$),
while the previous (Grassmann) treatment of Ref. \cite{Damour:2009zc} had assumed a linear coupling to the quadratic
spin operators. Let us also note that the link between our present dynamical reflection
operators, Eqs. \eqref{RNF2new}, \eqref{RNF3new}, and representations of $K[AE_3]$ is more indirect than what was
suggested by the Grassmann-based work of Ref. \cite{Damour:2009zc}. In particular, the 6-dimensional subspace
in which live the $N_F=2$ reflection operators is  strictly smaller than the full 15-dimensional $N_F=2$ 
space within which live the operators $\widehat J_{11}, \widehat S_{12}, \widehat S_{23}$ that carry
a representation of $K[AE_3]$. Moreover, the operators that appear in exponentiated form in Eqs. \eqref{RNF2new}, \eqref{RNF3new},
do not define a representation of $K[AE_3]$.

In view of our results, we can associate with the evolution of the supergravity state of the universe $|\Psi(\beta) \rangle$
(considered at each fermion level) a word in the group generated by the three reflection operators Eqs. \eqref{RNF2new}, \eqref{RNF3new},
i.e. a product of the form $ \cdots r_{i_n} r_{i_{n-1}} \cdots r_{i_2} r_{i_1}$. The matrix group generated by such
products is infinite. However, we must recall that our study was assuming a type of intermediate asymptotic behavior 
with a sparse sequence of wall collisions, separated by large enough distances in $\beta$ space to be able to treat
each collision of the wave packet as a separated one-wall reflection. Such an approximation is not expected to maintain
itself for an infinite number of collisions. Indeed, on the one hand, at level $N_F=3$ the (shifted) momentum $\pi'_a$ is timelike
($G^{ab} \pi'_a \pi'_b= - \frac12$) so that, after a finite number of reflections, one expects the trajectory of the wave packet 
to end up in a direction which does not meet anymore a potential wall. On the other hand, at level $N_F=2$ the (shifted) momentum 
$\pi'_a$ is spacelike ($G^{ab} \pi'_a \pi'_b= + \frac38$) so that, after a finite number of reflections, one expects
$\pi'_a$ to tip over, i.e. to migrate from the upper half [where $ \pi'^1+ \pi'^2+ \pi'^3 >0$, corresponding
to decreasing spatial volume ${\mathcal V}_3= abc=e^{-(\beta^1+\beta^2+\beta^3)}$] of its (one-sheeted) hyperboloidal mass-shell ,
to its lower half (corresponding to increasing spatial volumes). Such a cosmological bounce (further discussed in Ref. \cite{Damour:2014cba})
is then expected to generate a finite number of reflections during the re-expansion regime, before driving the wavefunction
in the (non-billiard-like) Friedman-type expansion regime. We leave to future work a discussion of the global evolution of the
quantum state of such a universe, which is classically expected to bounce back and forth, indefinitely, between large
volumes and small volumes (see Fig. 3 in Ref. \cite{Damour:2014cba}, and discussion in Sec. XX there).

\begin{acknowledgments}
We thank Igor Frenkel and Christophe Soul\'e for informative discussions.
Ph. S. thanks  IHES for its kind hospitality; his work has been partially supported by the PDR ``Gravity and extensions'' from 
the F.R.S.-FNRS (Belgium) (convention T.1025.14).
\end{acknowledgments}

\appendix

\section{Asymptotic plane-wave solutions} \label{appendixA}

We display hereafter the explicit values of the numerical constants entering the linear ($ L_{pq}^k \, \pi'_k$) and constant ($ m_{pq}$) 
contributions entering the amplitudes  
\begin{equation}
\label{Kpq}
K_{pq} \propto \pi'_p \, \pi'_q + L_{pq}^k \, \pi'_k + m_{pq}
\end{equation}
of the $N_F=2$ and $N_F=3$ asymptotic plane-wave solutions to which we referred in Eqs. \eqref{NF2wave}, \eqref{asNF3}.
For each wall form $\alpha^{\perp}$, with adapted basis $\alpha^{\widehat a}=\left\{ \alpha^{\perp}, \alpha^u, \alpha^v \right\}$, 
the values of the projected components $L^{\perp}_{\widehat a \widehat b}$ that vanish determine the global reflection phase factor
(see text).
\begin{itemize}
\item Level $N_F=2$:
\begin{equation}
L_{pq}^k \, \pi_k^\prime = -i \begin{pmatrix}
3 \, \pi'_1 + \pi'_2 + \pi'_3 &\frac32 (\pi'_1 + \pi'_2) &\frac12 (\pi'_1 + 3 \pi'_3) \\
{ \ } \\
\frac32 (\pi'_1 + \pi'_2) &2 \pi'_2 + \pi'_3 &\frac12 (\pi'_2 + \pi'_3) \\
{ \ } \\
\frac12 (\pi'_1 + 3\pi'_3) &\frac12 (\pi'_2 + \pi'_3) &\pi'_3
\end{pmatrix}
\quad,\quad
m_{pq} = -\frac14 \begin{pmatrix}
13 &9 &3 \\
9 &5 &1 \\
3 &1 &1
\end{pmatrix} \quad .
\end{equation}
\item Level $N_F=3$:
\begin{equation}\label{LkpqNF3CWa}
L_{pq}^k \, \pi_k^\prime = i \begin{pmatrix}
-\pi'_1 + \pi'_2 + \pi'_3 &-\frac12 \, \pi'_2 + \pi'_3 &-\frac12 (\pi'_1 - \pi'_3) \\
{ \ } \\
-\frac12 \, \pi'_2 + \pi'_3 &-\pi'_2 + \pi'_3 &-\frac12 \, \pi'_2  \\
{ \ } \\
-\frac12 (\pi'_1 - \pi'_3) &-\frac12 \, \pi'_2  &-\pi'_3
\end{pmatrix} 
\quad,\quad
m_{pq} = +\frac14 \begin{pmatrix} 5&2&1 \\ 2&2&0 \\1&0&-1 \end{pmatrix} \qquad .
\end{equation}
\end{itemize}
The above expressions correspond to the canonical billiard chamber $\beta^1<\beta^2<\beta^3$. See Ref. \cite{Damour:2014cba}
for a discussion of the other chambers.

\section{On finite-dimensional representations of $K[AE_3]$} \label{appendixB}

The finite-dimensional representations of the ``maximally compact'' subalgebra $K[AE_3]$ that naturally enter our
supergravity study constitute special ones. We have investigated more general finite-dimensional representations of 
$K[AE_3]$, and briefly report here some of our findings.

The algebra $K[AE_3]$ is defined as the subalgebra of the hyperbolic Kac--Moody $AE_3$ algebra \cite{Kac,FF} that is
fixed by the Chevalley involution $\omega$. We recall that the latter is defined by its action on the
Kac-Moody generators $(e_i,f_i,h_i)$: $\omega(e_i)=-f_i$, $\omega(f_i)=-e_i$ and $\omega(h_i)=-h_i$; so that,
for any Kac-Moody algebra ${\bm A}$, its maximally compact subalgebra $K[{\bm A}]$ is generated by the differences
$x_i \equiv e_i-f_i$. In the case of $AE_3$, with Dynkin diagram \eqref{dynkin}, this yields the three generators 
$x_1, x_2, x_3$, which are respectively equivalent (modulo a factor $i$) to the three generators $\widehat J_{11}$, $\widehat S_{12}$, and $\widehat S_{23}$.

Any three generators $\widehat J_{11}$, $\widehat S_{12}$, $\widehat S_{23}$  satisfying the following five relations \cite{Berman}\begin{equation}
\label{B1}
ad^2_{\widehat S_{23}} \, \widehat S_{12} =   \widehat S_{12} \, , \ ad^2_{\widehat S_{12}} \, \widehat S_{23} =   \widehat S_{23} \, ,
\end{equation}
\begin{equation}
\label{B2}
ad^3_{\widehat S_{12}} \, \widehat J_{11} =   4 \, ad_{\widehat S_{12}} \, \widehat J_{11} \,, \ ad^3_{\widehat J_{11}} \, \widehat S_{12} =  4 \, ad_{\widehat J_{11}} \, \widehat S_{12} \, ,
\end{equation}
\begin{equation}
\label{B3}
ad_{\widehat J_{11}} \, \widehat S_{23}=0 \, ,
\end{equation}
define a representation of  $K[AE_3]$. As in the text, we use here hermitian-type generators, corresponding to $-i x_1, -i x_2, -i x_3$
rather than antihermitian-type ones $x_i=e_i-f_i$, as generally used in mathematical works.

We are looking for finite-dimensional representations  of the three generators $\widehat J_{11}$, $\widehat S_{12}$, $\widehat S_{23}$
(i.e. three matrices, say $ J_{11}$, $ S_{12}$, $ S_{23}$), with emphasis on finding low-dimensional representations. [For a related study (oriented, however, towards finding high-dimension representations) in the case of $K[E_{10}]$ see Ref. \cite{Kleinschmidt:2013eka}.]
Conditions \eqref{B1} show that  $ S_{12}$ and $ S_{23}$ may be interpreted as usual $su(2)$ generators. Note that if,
given $ S_{12}$ and $ S_{23}$, a matrix   $ J_{11}$ satisfies relations \eqref{B2}, \eqref{B3}, so does the matrix 
$-  J_{11}$.  Moreover, the complex conjugate of any solution triplet of matrices $ J_{11}$, $ S_{12}$, $ S_{23}$ will also be a solution.  In addition, if  $ J_{11}$, $ S_{12}$, $S_{23}$ is a $n$-dimensional solution,
the triplet $J_{11} + k Id_n$, $ S_{12}$, $ S_{23}$ is also a solution for an arbitrary value of $k$.

One can first look for representations that are irreducible with respect to $su(2)$, i.e. with  $ S_{12}$ and $ S_{23}$
given, modulo conjugation, by the standard $(2j+1)$-dimensional, spin-$j$ matrices, say (with $m, m'$ varying by steps of 1 between $-j$ and $+j$)
\begin{eqnarray} \label{su2}
\left(S_{12}^{(j)}\right)_{m,\,m'}& =& \,m\,\delta_{m,\,m'}  \,,  \nonumber \\
\left(S_{23}^{(j)}\right)_{m,\,m'}& =&\frac 12 \left(\sqrt{(j-m)(j+m+1)}\,\delta_{m-1,\,m'}+\sqrt{(j+m)(j-m+1)}\,\delta_{m+1,\,m'} \right)\,.
\end{eqnarray}
The lowest-dimensional case would be the 2-dimensional spin-$\frac12$ $su(2)$ representation. However, we found that, in this case,
the only possible solutions of Eqs. \eqref{B2}, \eqref{B3} for $ J_{11}$ are $J_{11} \propto Id_2$.  In the present study, we consider
such solutions as being ``trivial"\footnote{We note, however, that the (real) 4-dimensional Dirac-spinor-type representation of $K[AE_3]$
discussed in Eqs. (4.14), (4.17) of  Ref. \cite{Damour:2009zc} is equal to a direct sum of two such  (complex-conjugated) 
2-dimensional representations with  ``trivial"  $ J_{11}$'s, and that the tensor product of the ``trivial" 2-dimensional representation,
and of the non-trivial 3-dimensional one discussed next, leads to a non-trivial 6-dimensional representation (see below).}.

The only $su(2)$-irreducible [cf. \eqref{su2}] representations with non-trivial $ J_{11}$ we found (up to $j=13/2$)
correspond   to $j=1$ and $j=\frac32$, i.e. to 3 and 4 dimensional representation spaces. We conjecture that these are the only such ones.

The lowest-dimensional nontrivial representation of $K[AE_3]$ is 3-dimensional. Its generators are given by \eqref{su2}  (for $j=1$),
together with
\begin{equation}\label{exJ1}
J^{(\pm1)}_{11} =   \pm \begin{pmatrix} k&0&1 \\ 0 &k+1&0\\1&0&k \end{pmatrix} \qquad ,
\end{equation}
whose eigenvalues are $\pm \,(k+1, k+1,\,k-1)$. Here, $k$, which corresponds to the shift $k Id_3$ mentioned above, is
arbitrary. One can choose $k= -\frac13$ if one wishes to normalize the trace of $J^{(\pm1)}_{11}$ to zero.

There is a similar 4-dimensional representation with generators given by \eqref{su2}  (for $j=\frac32$), together with
(modulo a $k Id$ shift)
\begin{equation}\label{exJ32}
J^{(\pm 3/2)}_{11} =   \pm \begin{pmatrix}
\frac12&0&- \frac{\sqrt{3}}2&0\\ 
0 &- \frac12&0&- \frac{\sqrt{3}}2\\
- \frac{\sqrt{3}}2&0&- \frac12&0\\
0&- \frac{\sqrt{3}}2&0& \frac12
 \end{pmatrix} \qquad ,
\end{equation}
whose eigenvalues are $\pm (1,\,1,\,-1,\,-1)$. 

Other kinds of representations exist, in which the spin generators are not irreducible. Actually, this is the case for the first-found
finite-dimensional representation of $K[AE_3]$, namely the 6-dimensional representation defined by the gravitino operators
in $3+1$-dimensional spacetime \cite{Damour:2009zc}. More precisely, Ref. \cite{Damour:2009zc} dealt with a {\it real}
12-dimensional representation, based on the transformations of a Majorana gravitino. However, it can easily be decomposed into two complex-conjugated 6-dimensional representations, each one of which is equivalent (modulo a suitable $ k Id_6$ shift of $J_{11}$)
to the complex, 6-dimensional representation of $K[AE_3]$ appearing at the $N_F=1$ level of our total quantum, 64-dimensional 
space. [The  6-dimensional $N_F=1$ representation we are talking about is the representation spanned by the six states $\tilde b^a_\pm | 0\rangle_-$.] In the latter representation, the spin operators are the direct sum of irreducible
representations with spins $\frac12$ and $\frac32$ [i.e. with eigenvalues of $S_{12}$ and $S_{23}$ equal to $ +\frac32, -\frac32,\left(+\frac12 \right)_2 \, , \left(-\frac12 \right)_2$].

Starting from 6-dimensional spin generators given by such a direct sum $(j=\frac12) \bigoplus (j=\frac32)$,  we looked for the most general $J_{11}$ 
satisfying the additional relations \eqref{B2}, \eqref{B3}. We found, in absence of additional requirements, multi-parameter families of solutions.
On the other hand, we can require that a non-degenerate sesquilinear form $H$ be left invariant by all the generators, i.e.
\begin{equation}\label{JHinv}
 J_{11}^\dagger \, H - H \, J_{11}=0 \,,
 \end{equation}
 and similar equations with the spin generators. [The relative minus sign in Eq. \eqref{JHinv} comes from the fact that, in our conventions,
 the one-dimensional group generated by $J_{11}$ is $e^{i \theta_{11} J_{11}}$.]
 The invariance of $H$ under the spin generators restricts it (in the basis of Eqs. \eqref{su2}) to the form
 $H=p\,Id_2\oplus q\, Id_4$, where (in the nondegenerate case) only the sign of the ratio $p/q$ matters.
 We then found that (besides isolated solutions) there exists four different  one-parameter families of such 6-dimensional representations.
 Parametrizing the elements of the 6-dimensional representation space as vector-spinors\footnote{Here, spinors mean two-component $su(2)$
 spinors $\xi^A$.} $v^{A \, a}$, with $A=1,2$  and $a=1,2,3$,  the generators $S_{12}, S_{23}, J_{11}(w)$ can be written in the factorized way\footnote{As in Ref. \cite{Damour:2009zc}, the presence of vectorial projectors $\alpha^a \alpha_b/\alpha^2$ in Eqs. \eqref{Jw} implies
 eigenvalues equal to $+3$ or $-1$ times those of the spinor matrices $\frac12 \sigma_i$.} discussed in Eqs. (3.11) and (4.16) of Ref. \cite{Damour:2009zc}, namely,
\begin{eqnarray} \label{Jw}
 \left( S_{12}\right)^{ A a}_{Bb} &=& \frac12 \left(\sigma_3\right)^A_B \left( 4 \, \frac{\alpha_{12}^a \,\alpha_{12 \, b}}{\alpha_{12} \cdot\alpha_{12 }} - \delta^a_{\, b} \right) \,, \nonumber \\
 \left(S_{23}\right)^{Aa}_{Bb} &=& \frac12 \left(\sigma_1\right)^A_B \left( 4 \, \frac{\alpha_{23}^a \,\alpha_{23 \, b}}{\alpha_{23} \cdot\alpha_{23 }} - \delta^a_{\, b} \right) \,, \nonumber \\
  \left(J_{11}^{E,L}(w)\right)^{Aa}_{Bb} &=& \frac12 \left(\sigma_0\right)^A_B \left( 4 \, \frac{\alpha_{w}^{a \, E,L} \,\alpha_{w \, b}}{\alpha_{w}^{E,L} \cdot\alpha_{w }} - \delta^a_{\, b} \right) \,. \
\end{eqnarray}
Here: $\sigma_3\equiv\sigma_z={\rm diag}(1,-1)$ and $\sigma_1\equiv\sigma_x$ are the usual (real) Pauli matrices; $\sigma_0\equiv Id_2$;
$\alpha_{12 \, a}$ and $\alpha_{23 \, a}$ are the same linear forms as in Eq. \eqref{defaij}; their contravariant versions $\alpha_{12}^a $,
$\alpha_{23}^a $ are defined by raising the index either by means of $G^{ab}$ or, equivalently, by means of $\delta^{ab}$; while the third,
gravitational-like linear form $\alpha_{w \, a}$ is the following one-parameter deformation of the usual gravitational linear form:
$\alpha_{w }(\beta)= \alpha_{w \, a} \beta^a= 2 \beta^1 + w (\beta^2+\beta^3)$. On the other hand, the third Eq. \eqref{Jw} involves,
depending on the value $E$ (for Euclidean) or $L$ (for Lorentzian) of the superscript, two different contravariant versions of $\alpha_{w }$,
namely, either $\alpha_{w}^{a \, E}  \equiv \delta^{ab} \alpha_{w \, b}$, or, $\alpha_{w}^{a \, L} \equiv G^{ab} \alpha_{w \, b}$, where $G^{ab}$
is the Lorentzian (contravariant) metric defined in Eq. \eqref{Gabu}. The parameter $w$ runs over the real line (except, in the Lorentzian case,
for the singular value $w=\frac12$ where the denominator 
$\alpha_{w}^{L} \cdot\alpha_{w } \equiv \alpha_{w}^{a \, L} \alpha_{w \, a} = 2 - 4w$ vanishes).
The eigenvalues of $J_{11}^{E,L}(w)$  depend neither on $w$ nor on the index $E, L$, and are equal to 
$ \left(+\frac32 \right)_2 \, , \left(-\frac12 \right)_4$. In addition to the two one-parameter families of 6-dimensional
representations displayed in Eqs. \eqref{Jw}, one can also define two other families obtained by changing the sign of
$J_{11}^{E,L}(w)$. When $w=0$ the Lorentzian solution Eqs. \eqref{Jw} is equivalent to the 6-dimensional representation inherited
from 4-dimensional supergravity discussed above (and appearing at $N_F=1$). On the other hand, when taking $w=0$
in the Euclidean solution Eqs. \eqref{Jw}, one gets (modulo a shift $k Id_6$ and a change in the sign of $J_{11}$)
the 6-dimensional representation obtained by taking 
the tensor product of the ``trivial" 2-dimensional representation discussed above (with
$J_{11} \propto Id_2$) with the 3-dimensional representation \eqref{exJ1} (say with $k=-\frac13$).
Finally, the components (with respect to the basis $v^{A \, a}$) of the invariant sesquilinear forms $H$, Eq. \eqref{JHinv}, 
of the representations \eqref{Jw} are, respectively, $\delta_{AB} \delta_{ab}$ for the Euclidean case, and
$\delta_{AB} G_{ab}$ for the Lorentzian case; $G_{ab}$ denoting the covariant form (i.e. the matrix inverse) of the
contravariant metric \eqref{Gabu}.


\end{document}